\documentclass{article}
\usepackage[utf8]{inputenc}
\usepackage[section]{placeins}
\usepackage{gensymb}
\usepackage{multicol}
\usepackage{graphicx}
\usepackage{framed}
\usepackage{setspace}
\usepackage{amsmath}
\usepackage{amssymb}
\usepackage{amsfonts}
\usepackage{mathrsfs}
\usepackage{mathtools}
\usepackage{array}
\usepackage[margin=1.25in]{geometry}
\usepackage{indentfirst}
\usepackage{amsopn}
 
\usepackage{float}
\usepackage{IEEEtrantools}
\usepackage{multirow}
\usepackage{hyperref}
\usepackage{cases}
\usepackage{caption}
\usepackage{listings}
\usepackage{stmaryrd}
\usepackage{setspace}
\usepackage[abbreviations = false]{siunitx}
\usepackage{wrapfig}
\usepackage{xfrac} 
\usepackage{array} 
\usepackage[text]{esdiff}
\usepackage{subcaption}

\usepackage{framed} 
\usepackage{pifont} 
\usepackage{enumerate} 
\usepackage{wasysym} 
\usepackage[super]{nth}

\usepackage{amsthm}  
\usepackage{authblk}

\theoremstyle{definition}

\theoremstyle{remark}

\usepackage[numbers]{natbib}
\usepackage{cite}
\usepackage{url}

\title{Investigating the influence of growth arrest mechanisms on tumour responses to radiotherapy}

\author[1]{Chloé Colson}
\author[1]{Philip K. Maini}
\author[1,2]{Helen M. Byrne}
\affil[1]{Wolfson Centre for Mathematical Biology, Mathematical Institute, University of Oxford, Radcliffe Observatory Quarter, OX2 6GG, Oxford, UK}
\affil[2]{Ludwig Institute for Cancer Research, Nuffield Department of Medicine, University of Oxford, Roosevelt Drive, Oxford, OX3 7DQ, UK}
\date{}                     
\begin{document}
\numberwithin{equation}{section}
 \numberwithin{rem}{section}

\maketitle

\begin{abstract}
Cancer is a heterogeneous disease and tumours of the same type can differ greatly at the genetic and phenotypic levels. Understanding how these differences impact sensitivity to treatment is an essential step towards patient-specific treatment design. In this paper, we investigate how two different mechanisms for growth control may affect tumour cell responses to fractionated radiotherapy (RT) by extending an existing ordinary differential equation model of tumour growth. In the absence of treatment, this model distinguishes between growth arrest due to nutrient insufficiency and competition for space and exhibits three growth regimes: nutrient-limited (NL), space limited (SL) and bistable (BS), where both mechanisms for growth arrest coexist. We study the effect of RT for tumours in each regime, finding that tumours in the SL regime typically respond best to RT, while tumours in the BS regime typically respond worst to RT.  For tumours in each regime, we also identify the biological processes that may explain positive and negative treatment outcomes and the dosing regimen which maximises the reduction in tumour burden.
\end{abstract}


\section{Introduction}\label{sec1}
Understanding the biological mechanisms underpinning cancer and developing effective therapeutic protocols to improve patient prognosis are fundamental aims of cancer research. Existing treatment modalities, such as radiotherapy (RT) and chemotherapy (CT), are applied via highly-regulated dosing protocols \citep{RT,CT} to avoid damaging healthy tissue, while maximising treatment effect. Nonetheless, the efficacy of both RT and CT is limited by their intolerable side-effects. Further, inter-tumour heterogeneity can significantly influence sensitivity to treatment.  Investigating how different growth mechanisms may affect response to treatment is, therefore, an important step towards overcoming barriers to treatment efficacy. In this paper, we investigate how two distinct growth-rate limiting mechanisms, namely growth arrest in response to nutrient insufficiency and to competition for space, impact tumour response to RT. 

\paragraph{A dynamic model of tumour growth that distinguishes between mechanisms of tumour control.} 
Regardless of their biological complexity, existing models of solid tumour growth typically describe a single mechanism by which a tumour may reach an equilibrium.  For example, the models developed by \citep{enderling2006mathematical,liu2021time,milzman2021modeling,zahid2021dynamics} predict growth arrest due to a cessation of proliferation (with no explicit cell death), while those proposed by \citep{drasdo2005single,greenspan1972models,hillen2013tumor,lewin2020three} predict growth arrest due to the balance of cell proliferation and cell death. 

In previous work \citep{colson2022combining}, we developed a model of tumour growth that simultaneously describes growth arrest due to nutrient inhibition, when the net rates of cell proliferation and death are equal (and strictly positive), and growth arrest due to contact inhibition, when the net rate of cell proliferation becomes zero, with no cell death. We assumed that the system is well-mixed and neglected angiogenesis and vascular remodelling. In more detail, we viewed the vascular volume as a parameter which influences nutrient and space availability and, therefore, the tumour's carrying capacity. The model comprises two time-dependent ordinary differential equations (ODEs) for the tumour volume, $T(t)$, and the oxygen concentration, $c(t)$, and can be written as follows:
\begin{equation}
\diff{T}{t} = \underbrace{q^*_2cT(S_{max} - (T+V_0))}_\text{\shortstack{rate of \\ tumour cell proliferation}} - \underbrace{\left[\delta^*_1(c^*_{min} - c)\right] H(c^*_{min}-c)T}_\text{\shortstack{rate of \\ cell death due to nutrient starvation}},
\label{eq:1.1}
\end{equation}
\begin{equation}
\diff{c}{t} = \underbrace{g^*(c^*_{max} -c)V_0}_\text{\shortstack{rate of \\ nutrient delivery}} - \underbrace{q^*_1cT}_\text{\shortstack{baseline rate of \\ nutrient consumption} }- \underbrace{q^*_3cT(S_{max} - (T+V_0))}_\text{\shortstack{additional rate of \\ nutrient consumption for proliferation} },
\label{eq:1.2}
\end{equation}
where
\begin{equation}
H(x) = \begin{cases}
1, \,\, \text{if } x \geq 0, \\ 0, \,\, \text{if } x < 0.
\end{cases}
\label{eq:1.3}
\end{equation}
Denoting the total available space by $S_{max}  \, (m^3)$ and the vascular volume by $V_0 \, (m^3)$, the rate of tumour cell proliferation is assumed to be proportional to the available free space, $S_{max}-T-V_0$, and to the oxygen concentration, $c$, with proportionality constant $q_2^*\,(\si{\kilogram}^{-1} \si{\minute}^{-1} )$.  If $c$ drops below a threshold value, $c^*_{min} \, (kg/m^3)$, then cells die at a rate proportional to the difference between $c$ and $c^*_{min}$, with proportionality constant $\delta_1^* \, (\si{\metre \cubed} \si{\kilogram}^{-1} \si{\minute}^{-1} )$.  Further, oxygen is supplied to the tumour at a rate proportional to $V_0$ and the difference between the oxygen concentration in the vasculature,  $c^*_{max} \, (kg/m^3)$, and in the tumour.  The parameter $g^*$ is the rate of oxygen exchange per unit volume area of blood vessel.  Finally, oxygen is consumed by tumour cells for maintenance at a rate proportional to $c$, with rate constant $q_1^* \, (\si{\metre}^{-3}  \si{\minute}^{-1} )$, and for proliferation at a rate proportional to the proliferation rate, with conversion factor $k > 0$ defined such that $q_3^* =  q_2^*/k \, (\si{\metre}^{-6}  \si{\minute}^{-1}) $.

Since the model \eqref{eq:1.1}-\eqref{eq:1.2} distinguishes between two mechanisms for growth-control, it can be used to investigate how they impact tumour response to treatment.  Therefore, in this work, we extend Equations \eqref{eq:1.1}-\eqref{eq:1.2} to account for the biological effects of RT.

\paragraph{Radiobiology.} 
RT is used to treat more than $50\%$ of cancer patients \citep{maier2016cellular}. It involves the delivery of energy rays, via small doses called fractions, at regular time intervals and over a fixed period of time, to a region of the body comprising both cancerous and healthy tissue. Radiation protocols are, therefore, designed to balance treatment efficacy and undesirable side-effects in normal tissues. While a conventional fractionation schedule consists of $2 \, \mathrm{Gy}$ doses delivered Monday to Friday for up to $7$ weeks \citep{ahmed2014altered}, the dose, dosing frequency and treatment duration can be varied to deliver a fixed total dose. The latter is often termed as the Maximum Tolerated Dose (MTD), i.e., the highest dose which does not cause adverse side-effects \citep{GAD2014164}.  

Radiation induces direct and indirect cytotoxic effects by causing DNA damage to cancer cells that is fatal if left unrepaired.  Direct effects arise from interactions between ionising particles and DNA and indirect ones from interactions between ionising particles and water, which create reactive oxygen species that subsequently react with DNA. Indirect effects are the most common, which is why hypoxic, i.e., poorly oxygenated, tumours are often radio-resistant \citep{graham2018overcoming}.  Intratumoural oxygen levels are a key factor influencing tumour radio-sensitivity, another being tumour cell proliferation rates as cells that are in the G2 or mitosis phases of the cell cycle are the most sensitive to RT.  

RT can also affect the tumour vasculature, with increases in angiogenesis observed at low doses \citep{marques2020low} and vascular damage and necrosis observed at high doses \citep{stolz2022multiscale, venkatesulu2018radiation}.  In this work, we neglect the effect of RT on vasculature in order to focus on evaluating how nutrient- and contact-inhibited growth control affect the sensitivity of tumour cells to treatment with RT.

\paragraph{Mathematical modelling of tumour response to radiotherapy.}
A number of mathematical models have been proposed to describe tumour response to RT. Key aims of these modelling efforts include studying specific RT protocols \citep{enderling2010quantitative,jeong2017modeling, lewin2018evolution, prokopiou2015proliferation,rockne2009mathematical}, designing patient-specific RT dosing schedules \citep{alfonso9estimating,belfatto2018model} and investigating the influence of inter- and/or intra- tumour heterogeneity on tumour sensitivity to RT \citep{alfonso2019modeling, celora2023spatio,enderling2009importance,powathil2012modeling,watanabe2016mathematical}. 

While the purpose of these modelling approaches may differ, they are all based on the common assumption that RT inflicts instantaneous cell death on tumour cells and the cell kill is modelled using the Linear-Quadratic (LQ) model \citep{mcmahon2018linear}. The LQ model states that the fraction, $S_{LQ}$, of (tumour) cells that survive exposure to a dose $D$ ($\mathrm{Gy}$) of radiation is given by
\begin{equation}
S_{LQ}(D) = \exp\left( - \left(\alpha D + \beta D^2\right)\right),
\label{eq:1.4}
\end{equation}
where $\alpha \geq 0 $ and $\beta \geq 0$ are tissue-specific radio-sensitivity parameters.  These parameters are typically derived from cell survival data collected at a small number of time points in \textit{in vitro} 2D monolayer or 3D spheroid experiments.  As such, they provide information about the long-term proportion of cell death rather than how the cell death rate changes over time.  In contrast,  a time-dependent description of RT cell kill can account for different types of damage (direct vs. indirect), damage repair and cell death following insufficient repair \citep{curtis1986lethal,goodhead1985saturable, neira2020kinetic, tobias1985repair}. Such a description facilitates the study of the evolution of tumour composition during treatment, as we may keep track of changes in healthy, damaged and dead tumour cell populations.  In this paper, we follow \citet{neira2020kinetic} and adopt a time-dependent description of RT.

\paragraph{Paper structure.}
This paper is structured as follows. In Section \ref{sec:2}, we extend the tumour growth model defined by Equations \eqref{eq:1.1}-\eqref{eq:1.2} to account for the biological effects of RT.  We summarise the key features of the model dynamics in the absence of treatment in Section \ref{sec:3}. Then, we investigate the response of tumours characterised by different growth-limiting mechanisms in Section \ref{sec:4}, initially performing a numerical study of tumour response during RT and, subsequently, looking at post-treatment growth dynamics via a steady state analysis and complementary numerical study.  The paper concludes in Section \ref{sec:5}, where we discuss our findings and outline possible avenues for future work.

\section{Model development} \label{sec:2}
\subsection{The mathematical model}
In this section, we incorporate tumour response to RT into the growth model \eqref{eq:1.1}-\eqref{eq:1.2}. We follow the approach outlined in \citet{neira2020kinetic} and adopt a time-dependent description of radiotherapy. In more detail, we introduce the dependent variables $T_S$ and $T_R$ to denote tumour cells that have been sub-lethally and lethally damaged by RT. We suppose that the tumour is exposed to a total dose $D \, (\mathrm{Gy})$ of RT at a constant rate $R$ over the time period $t_{R} \leq t \leq t_{R}+\delta_{R} \, (\si{\min})$ so that
\begin{equation}
R(t) = \begin{cases} D/\delta_{R}, \quad \text{if} \, \, t_{R} \leq t \leq t_{R}+\delta_{R}, \\
0, \qquad\quad  \text{otherwise.}\end{cases}
\label{eq:2.1}
\end{equation}

\begin{figure}[!h]
\centering
\includegraphics[scale=0.4]{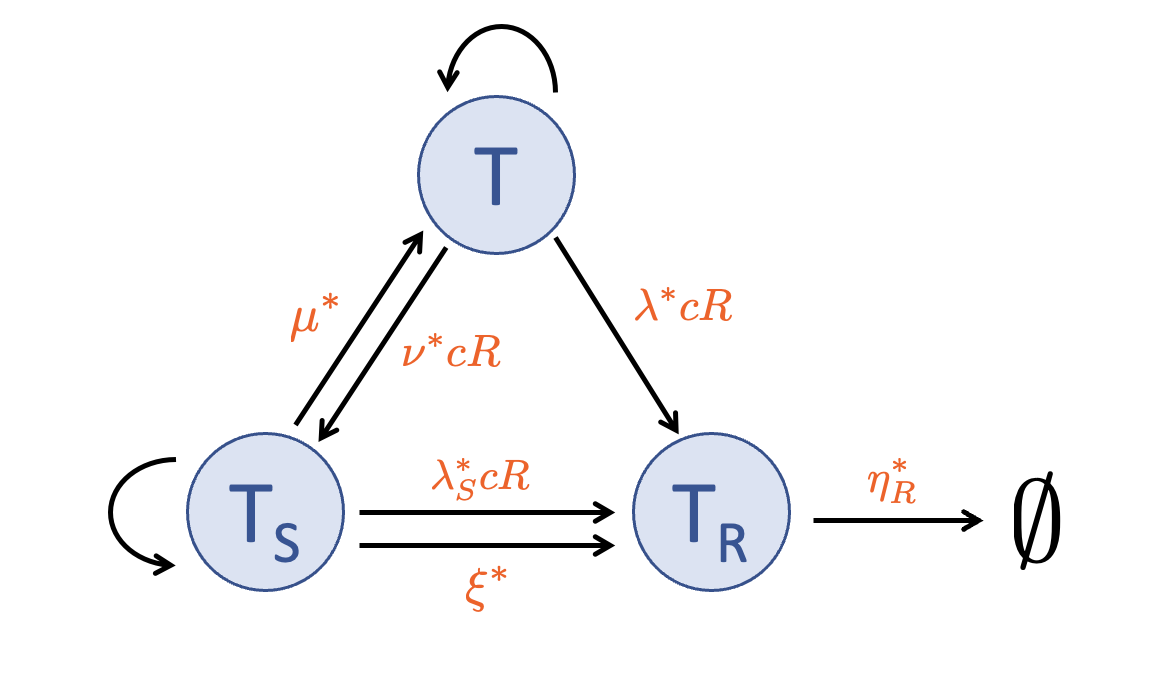} 
\caption{Schematic showing the interactions between undamaged, damaged and dead tumours cells, $T$, $T_S$, $T_R$, respectively,  in response to RT and the proliferation of $T$ and $T_S$ cells as described in the model defined by Equations \eqref{eq:2.1}-\eqref{eq:2.5}.  $R$ denotes the RT dose rate defined by \eqref{eq:2.1} and $c$ denotes the intratumoural oxygen concentration. }
\label{fig:2.1}
\end{figure}

Let $\Sigma = T+T_S+T_R+V_0$ be the total tumour volume.  We propose the following system of time-dependent ODEs to describe tumour growth and response to RT (see also the schematic in Figure \ref{fig:2.1}):
\begin{multline}
\diff{T}{t} = {q_2^*}cT(S_{max} - \Sigma) - \delta^*_1(c^*_{min} - c)H(c^*_{min}-c)T \\ { - \underbrace{ \lambda^* c R T}_\text{\shortstack{rate of direct \\ lethal damage}} - \underbrace{\nu^* c R T}_\text{\shortstack{rate of \\ sub-lethal damage}} + \underbrace{\mu^* T_{S}}_\text{\shortstack{rate of \\ repair}}}, 
 \label{eq:2.2}
\end{multline}

\begin{multline}
{\diff{T_{S}}{t} = { q^*_{2,S}}cT_{S}(S_{max} -\Sigma) - \delta^*_{1,S}(c^*_{min} - c)H(c^*_{min}-c)T_{S}}\\ {+{\nu^* c RT}-{\mu^* T_{S}} - \underbrace{\xi^* T_{S}}_\text{\shortstack{rate of \\ post-RT death \\ due to MC} }-\underbrace{\lambda^*_{S} c R  T_{S}}_\text{\shortstack{rate of indirect \\ lethal damage}}} , 
 \label{eq:2.3}
\end{multline}
\begin{equation}
{\diff{T_{R}}{t} ={\lambda^* c R  T} +{(\xi^*+\lambda^*_{S}c R)  T_{S}}- \underbrace{\eta^*_R T_{R}}_\text{\shortstack{rate of \\ clearance}}}, 
\label{eq:2.4}
\end{equation}
\begin{multline}
\diff{c}{t} = g^*(c_{max} -c)V_0 - q^*_1 c T - q^*_3 c T(S_{max} - \Sigma) { -  q^*_{1,S} c T_{S}  - q^*_{3,{S}} c T_{S} (S_{max} - \Sigma)},
\label{eq:2.5} 
\end{multline}
where $H$ is the Heaviside function defined in \eqref{eq:1.3}.

We assume that undamaged tumour cells, $T$,  proliferate, die due to nutrient insufficiency and consume oxygen for proliferation and maintenance as in Equations \eqref{eq:1.1}-\eqref{eq:1.2}. They suffer sub-lethal and lethal damage during irradiation at rates proportional to the oxygen concentration, $c$, and the RT dose rate, $R$, with proportionality constants $\nu^* > 0$ and $\lambda^* > 0$, respectively. We further suppose that sub-lethal damage is either repaired at a constant rate $\mu^* > 0$, or leads to tumour cell death via two distinct pathways.  First, sub-lethal damage may become lethal as it accumulates at a rate proportional to the oxygen concentration, $c$, and the RT dose rate, $R$, with proportionality constant $\lambda^*_{S}> 0$.  Second, sub-lethally damaged cells, $T_S$, may also undergo mitotic catastrophe (MC) if they attempt to divide with mis- or un-repaired DNA damage; we assume this occurs at a constant rate $\xi^*>0$. 

Sub-lethally damaged cells, $T_{S}$, also consume oxygen for maintenance and proliferation, proliferate and die due to nutrient insufficiency similarly to undamaged cells, $T$, although at different rates. More specifically,  they proliferate at a rate proportional to the oxygen concentration, $c$, and the available space, with proportionality constant $q^*_{2,S} = \theta_2 q^*_2 > 0$, with $\theta_2 \in (0,1)$. The latter ensures that damaged cells proliferate more slowly than undamaged cells as they expend more energy repairing RT damage than proliferating. Accordingly, sub-lethally damaged cells consume oxygen for maintenance at a rate proportional to $c$, with rate constant $q^*_{1,S} = \theta_1 q^*_1 > 0$, where $\theta_1 > 1$ as these cells require more energy to repair RT damage. They also consume oxygen for proliferation at a rate proportional to the proliferation rate, with conversion factor $k > 0$ such that $q^*_{3,S} = \frac{q^*_{2,S}}{k}$.  Here, we assume the same conversion factor for $T$ and $T_S$ cells, for simplicity.  Since $q^*_{2,S} = \theta_2 q^*_2 $ and $q^*_{3} = \frac{q^*_{2}}{k}$, we also have $q^*_{3,S} = \theta_2 q^*_3 $, i.e., damaged cells consume less oxygen for proliferation than undamaged cells. Lastly, as for $T$ cells, $T_S$ cells die from nutrient insufficiency when $c < c^*_{min}$, at a rate proportional to the difference between $c$ and $c^*_{min}$, with proportionality constant $\delta^*_{1,S} > 0$.  

Lethally-damaged cells, $T_{R}$, are considered to be dead: their damage cannot be repaired, they do not consume oxygen or proliferate, but they occupy space and are degraded at a constant rate $\eta^*_R > 0$.  

One final and important assumption we make is that radiation only affects tumour cells, i.e., we neglect any effects RT may have on tumour angiogenesis, vascular remodelling and injury.  This simplifying assumption enables us to focus on elucidating how mechanisms of growth arrest influence one particular type of tumour response to RT, i.e. the cellular response.

\subsection{Non-dimensionalisation}
We non-dimensionalise Equations \eqref{eq:2.1}-\eqref{eq:2.5} by introducing the following scalings:
\begin{align*}
& \widehat{T} = \frac{T}{S_{max}}, \quad \widehat{T}_{S} = \frac{T_{S}}{S_{max}}, \quad \widehat{T}_{R} = \frac{T_{R}}{S_{max}},\quad \widehat{V_0}=\frac{V_0}{S_{max}},  \\& \widehat{c}= \frac{c}{c_{max}}, \quad \widehat{R}= \frac{R}{R_{max}}, \quad \widehat{t} =  \frac{t}{\tau}.
\end{align*}
The timescale of interest is fixed to be $\tau = 1 \, \si{\min}$, which is the timescale for the duration of RT, and the maximum dose rate $R_{max} = 1 \, \mathrm{Gy}/\mathrm{min}$ \citep{konopacka2016can}. Then, given that  $q^*_{1,S} = \theta_1 q^*_1$, $q^*_{2,S} = \theta_2 q^*_2$  and $q^*_{3,S} = \theta_2 q^*_3$ and dropping hats for notational convenience, we obtain the following dimensionless system:
\begin{multline}
\diff{T}{t} = q_2 cT(1 - \Sigma) - (\delta_1(c_{min} - c)H(c_{min}-c) + \lambda c R + \nu c R)T + \mu T_{S},
 \label{eq:2.7}
\end{multline}
\begin{multline}
\diff{T_{S}}{t} = {\theta_2 q_2 }cT_{S}(1 - \Sigma) - (\delta_{1,S}(c_{min} - c)H(c_{min}-c) + \lambda_{S} c R + \mu +\xi )T_S +{\nu c R T},
 \label{eq:2.8}
\end{multline}
\begin{equation}
\diff{T_{R}}{t} ={\lambda c R T} +{(\xi+\lambda_{S} c R)  T_{S}}- \eta_R T_{R},
\label{eq:2.9}
\end{equation}
\begin{equation}
{\diff{c}{t} =g(1-c)V_0 - q_1 (T + \theta_1 T_S) c  - q_3 \left(T+\theta_2 T_S\right)c(1- \Sigma),}  
\label{eq:2.10} 
\end{equation}
where we have introduced the following dimensionless parameter groupings:
\begin{equation}
\begin{aligned}
 & {q_1}= {q^*_1 S_{max} \tau}, \; {q_3}={q^*_3 S_{max}^2 \tau}, \; {q_2}={q^*_2 S_{max}^2 c^*_{max} \tau}, \; k = \frac{c^*_{max}}{S_{max}}k^*, \;  \\
 & c_{min} = \frac{c^*_{min}}{c^*_{max}}, \; \delta_1= {\delta_1^* c_{max}\tau}, \; \delta_{1,S}= {\delta_{1,S}^* c_{max} \tau}, \; g = {g^* S_{max}\tau}, \\
 & \lambda= {\lambda^* c_{max} R_{max} \tau}, \; \lambda_S= {\lambda_S^* c_{max} R_{max} \tau}, \; \nu= {\nu^* c_{max} R_{max} \tau}, \\
 & \mu={\mu^* \tau}, \; \xi = {\xi^* \tau},  \; \eta_R = {\eta_R^* \tau}.
\end{aligned}
\label{eq:2.11}
\end{equation}

\subsection{Defining the dimensionless model parameters}
This paper focusses on studying the impact of two distinct growth arrest mechanisms on the qualitative tumour response to RT. We, therefore, fix parameters related to tumour cell responses to RT at the values stated in Table \ref{tab:3.1}. The values of $\nu$, $\lambda$, $\lambda_S$, $\mu$ and $\eta_R$ are based on values found in the literature \citep{neira2020kinetic, steel1987dose} and we assume $\lambda = \lambda_S$, for simplicity. We also set $\xi = 5 \times 10^{-4}$ so that cells that undergo mitotic catastrophe have a half-life of approximately $24\mathrm{h}$. Here, we implicitly assume that the average duration of the cell cycle in healthy cells \citep{bernard2006cells} and cancer cells are approximately the same. The parameters that define the RT dosing schedules (e.g. the dose rate, $R$) are summarised in the Methods section. 

Further, we define tumour growth parameters as in our previous work \citep{colson2022combining}, with the additional simplifying assumption that, as for the undamaged cells, $q_{1,S} = q_{2,S}$. We also set $\theta_1 = 10$ and $\theta_2 = 0.1$ to represent a $10$-fold increase in oxygen consumption for maintenance and a $10$-fold decrease in oxygen consumption for proliferation in damaged cells.

\begin{table}[ht!]
\renewcommand{\arraystretch}{1.5} 
\centering
\begin{tabular}{ c | c | c }
\hline
Parameters &  Definition & Value(s)  \\ \hline \hline
 $c_{min}$ & Anoxic oxygen threshold &  $10^{-2}$  \\ 
 $g$ & Rate of oxygen exchange per unit vascular volume & $5$ \\ 
 $k$ & Conversation factor &$10^{-2}$ \\ 
$q_1$ &  $O_2$ consumption rate for maintenance  &$  [10^{-2}, 10]$\\
$\theta_1$ & Proportionality constant relating $q_1$ and $q_{1,S}$& $ 10$\\
$q_3$& $O_2$ consumption rate for proliferation   &$ [10^{-2}, 10]$\\ 
$q_{2} $ & Proliferation rate &$  q_{3}/k$\\ 
$\theta_2$ & Proportionality constant relating $q_2$ and $q_{2,S}$ ($q_3$ and $q_{3,S}$) & $ 0.1$\\ 
$\delta_1,\delta_{1,S}$ & Rates of death due to nutrient insufficiency & $q_2, \theta_2 q_2 $ \\
$V_0$ & Vascular volume  & $(0,5 \times 10^{-3}]$  \\ 
$\nu$ & RT sub-lethal damage rate & $ 10$  \\
$\lambda, \lambda_{S}$ & RT lethal damage rate & $ 1$   \\
$\mu$ & Repair rate constant & $ 5\times 10^{-3}$  \\ 
$\zeta$ & Rate of death by mitotic catastrophe & $ 5 \times 10^{-4}$ \\ 
$\eta_R$ & Clearance rate of cells killed by RT & $  5 \times 10^{-5}$  \\ \hline
\end{tabular}
\caption{List of dimensionless parameters and their default values.}
\label{tab:3.1}
\end{table}

\section{Review of the key model dynamics in the absence of treatment} \label{sec:3}
\begin{figure}[ht!]
\centering
\includegraphics[scale=0.4]{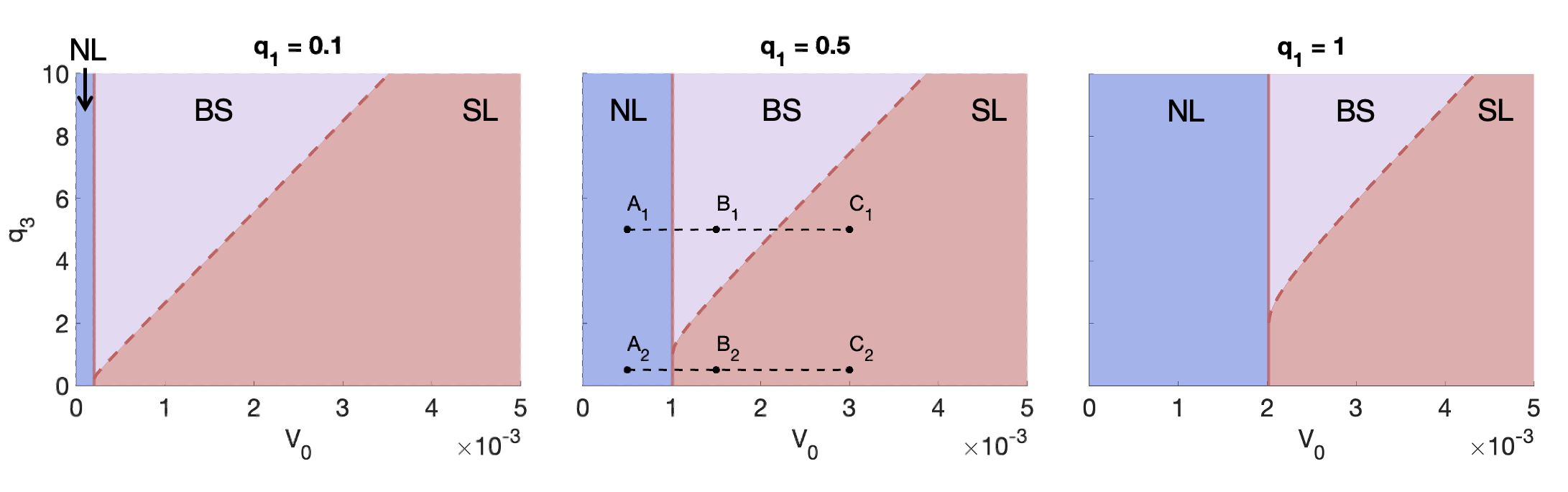} 
\caption{For $q_1 \in \{0.1, 0.5, 1\}$, we show the regions of $(V_0, q_3)$-space in which only the stable NL steady state exists (blue),  only the stable SL steady state exists (red) and both the stable NL and SL steady states co-exist (purple).  The solid and dashed red lines represent the boundaries between the three regions. For $q_1 = 0.5$, the points $A_1$-$C_1$ correspond to $(V_0,q_3) \in  \{(0.0005,5), (0.0015,5), (0.003,5) \}$, respectively, and $A_2$-$C_2$ to $(V_0,q_3) \in  \{(0.0005,0.5), (0.0015,0.5), (0.003,0.5) \}$, respectively.  Tumours defined by parameter sets $A_1$-$C_1$ have values of $q_3$ which are sufficiently larger than $q_1$ that there is bi-stability, while, for $A_2$-$C_2$, bistability does not occur.}
\label{fig:3.1}
\end{figure}

In this section, we summarise the model behaviour in the absence of treatment. Setting $R \equiv 0$ in Equations \eqref{eq:2.7}-\eqref{eq:2.10}, we recover the dimensionless form of Equations \eqref{eq:1.1}-\eqref{eq:1.2}. In \citep{colson2022combining}, we showed that this model admits two non-trivial, stable steady states (see Appendix A):
\begin{enumerate}
\item a nutrient limited (NL) steady state, attained when cell proliferation balances cell death due to nutrient starvation;
\item a space limited (SL) steady state, attained when cell proliferation ceases due to lack of space, with no cell death.
\end{enumerate}
For these solutions to be physically realistic and lie in the appropriate nutrient regime, they must satisfy $0 \leq T < 1-V_0$ and either $0 \leq c < c_{min}$, for the NL steady state, or $ c \geq c_{min}$, for the SL steady state. Imposing these conditions, we find that admissible NL and SL steady state solutions lie in different regions of parameter space. For fixed values of $c_{min} = 0.01$, $g = 5$ and $k = 0.01$, Figure \ref{fig:3.1} shows these regions in $(q_3,V_0)$-space for three values of $q_1$.

Given $q_1$, there exists a threshold value of $V_0$, say $V_N$, which is independent of $q_3$, such that only the NL steady state exists for $0 < V_0 \leq V_N$. Tumours in this region of parameter space, e.g. $A_1$ and $A_2$, are said to be in an NL growth regime. Further, given $q_1$ and $q_3$ sufficiently large relative to $q_1$, there exists another threshold value of $V_0$, say $V_S$, such that only the SL steady state exists for $V_0 \geq V_S > V_N$.Tumours in this region of parameter space, e.g. $C_1$, $B_2$ and $C_2$ are said to be in an SL growth regime. In this case, for $V_N < V_0 < V_S$, the NL and SL steady states co-exist.  Thus, in this region of parameter space,  a tumour, e.g. $B_1$, may evolve to either steady state depending on its initial conditions. We consider such tumours to be in a bistable (BS) growth regime. Finally, for $q_3 \lesssim q_1$, we have that, for $V_0 > V_N$, a unique steady state exists and it is of SL type.

\begin{figure}[!h]
\begin{subfigure}[!htb]{0.5\textwidth}
\centering
\includegraphics[scale=0.3]{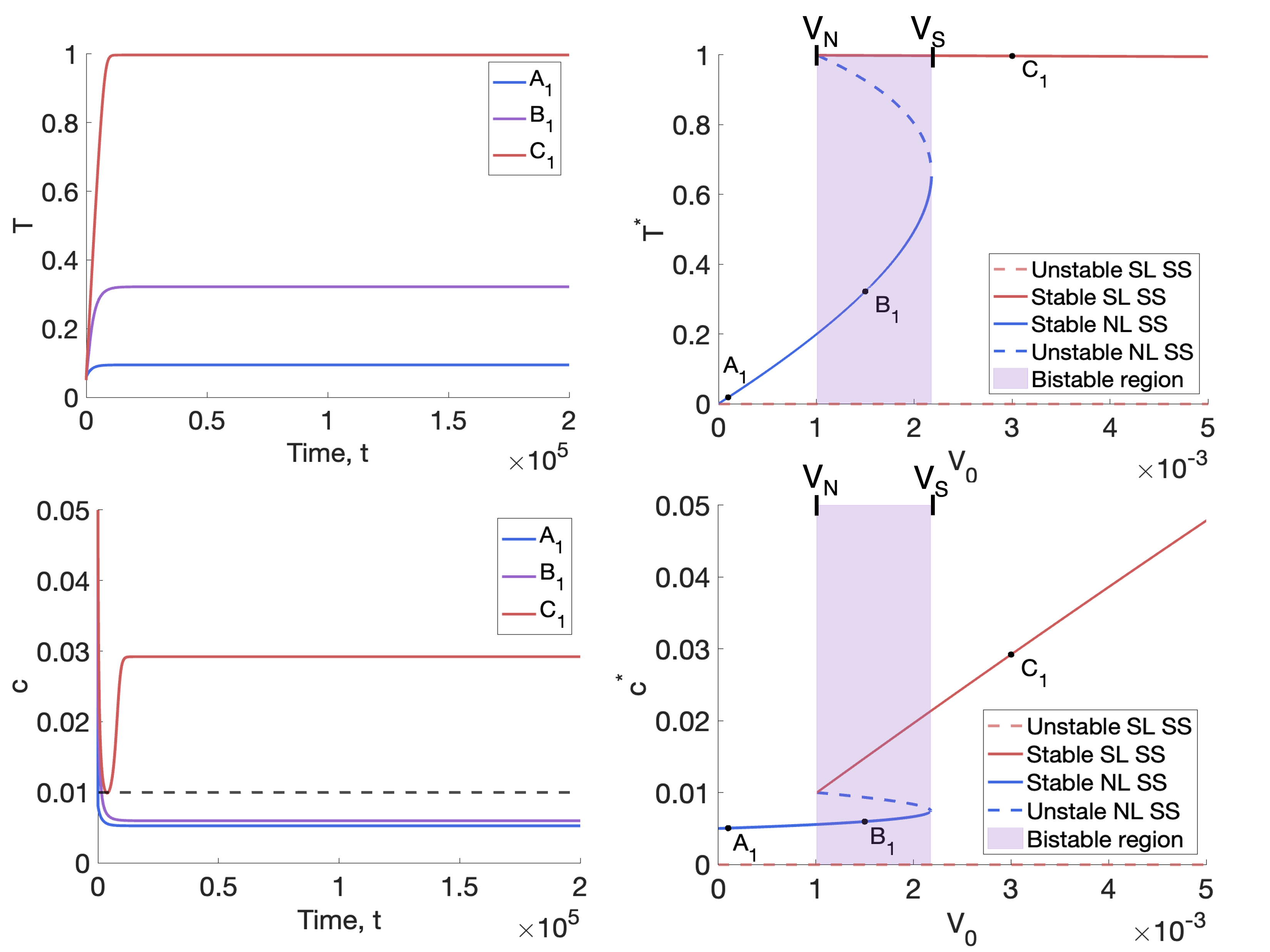} 
\caption{}
\label{fig:3.2A}
\end{subfigure}
\hfill
\begin{subfigure}[!htb]{0.5\textwidth}
\centering
\includegraphics[scale=0.3]{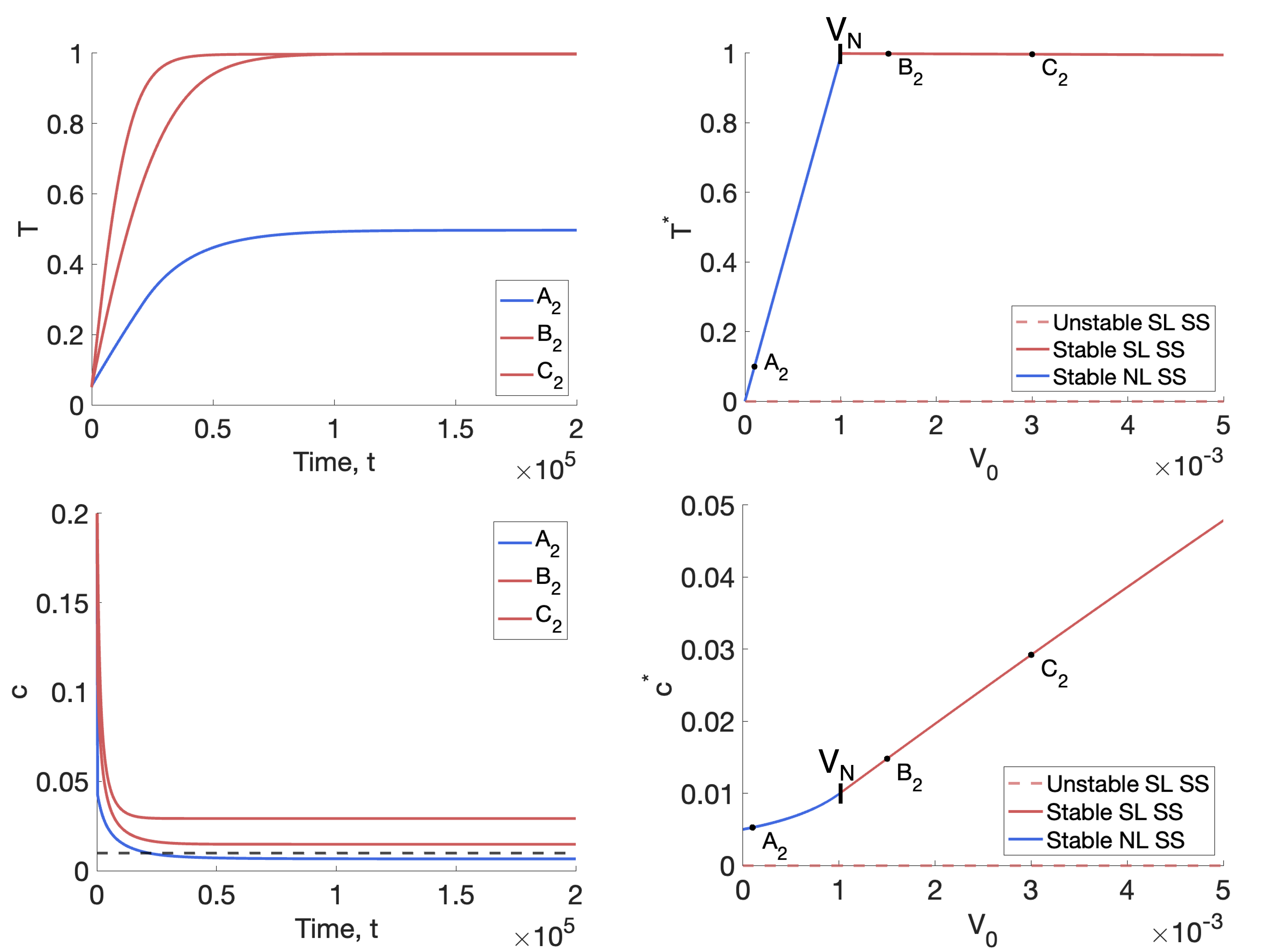} 
\caption{}
\label{fig:3.2B}
\end{subfigure}
\caption{With $R \equiv 0$, we numerically solve Equations \eqref{eq:2.7}-\eqref{eq:2.10} for $t \in (0,2 \times 10^5]$ subject to the initial conditions $(T(0), T_S(0), T_R(0),c(0)) = (0.05, 0,0,1)$ and plot the evolution of the tumour volume and oxygen concentration in time. In (a), $(q_1,q_3,V_0)$ correspond to points $A_1$-$C_1$ in Figure \ref{fig:3.1} and, in (b), they correspond to points $A_2$-$C_2$. The remaining model parameters are fixed at the default values in Table \ref{tab:3.1}.  The bifurcation diagrams show how the steady state values for $T$ and $c$ change as $V_0$ varies for $(q_1,q_3)$ corresponding to $A_1$-$C_1$ in (a) and for $(q_1,q_3)$ corresponding to $A_2$-$C_2$ in (b).  In both cases, the NL steady state increases with $V_0$ and is smaller than the SL steady state, which decreases with $V_0$. The tumour in a BS regime ($B_1$) grows to its NL steady state. }
\label{fig:3.2}
\end{figure}

Figure \ref{fig:3.2} shows the time evolution of the tumour cell volume, $T$, and the intratumoural oxygen concentration, $c$, and the corresponding bifurcation diagrams for tumours $A_1$-$C_1$ and $A_2$-$C_2$. In all cases, the NL steady state values for $T$ and $c$ ($T^*$ and $c^*$) are smaller than the SL ones. This is consistent with the assumption that, in the absence of angiogenesis, well-oxygenated tumours attain larger volumes than poorly-oxygenated tumours.  Further,  tumours in a BS regime evolve to their NL steady state for initial conditions satisfying $ 0 < T(0) \ll T^*$, which we use to simulate tumour growth. As the values of $T^*$ and $c^*$ for NL tumours increase with $V_0$, tumours in a BS regime will grow to larger volumes than tumours in an NL regime (and smaller volumes than tumours in a SL regime).  We also note that, in BS regimes, there is a large jump in $T^*$ and $c^*$ at $V_S$, the threshold value of $V_0$ separating the BS and SL regimes. In contrast, in monostable regimes, $T^*$ and $c^*$ depend continuously on $V_0$.

In summary, in the absence of treatment (i.e. $R \equiv 0$), Equations \eqref{eq:2.7}-\eqref{eq:2.10} describe three possible scenarios for tumour growth: (i) nutrient limited growth, where the tumour grows to an NL steady state at which proliferation balances cell death due to nutrient insufficiency; (ii) space limited growth, where the tumour grows to a SL steady state at which proliferation ceases due to space constraints; (iii) bistable growth, where a tumour grows to a NL steady state given physically realistic initial conditions ($0 < T(0) \ll T^*$).  In Section \ref{sec:4}, we investigate how tumours in these growth regimes respond to RT in order to understand how different growth arrest mechanisms may influence tumour response.

\section{Investigating tumour response to radiotherapy} \label{sec:4}
\subsection{Methods}
Our aim is to understand the qualitative response of tumours in nutrient limited (NL), space limited (SL) and bistable (BS) regimes to a range of fractionated radiotherapy (RT) treatments.  As a first step, we create three virtual tumour populations as follows. We first fix all tumour growth model parameters, except $q_1$, $q_3$ and $V_0$, at the default values stated in Table \ref{tab:3.1}. We then also fix $V_0 = 0.0005$, $V_0 = 0.005$ and $V_0 = 0.00275$ for tumours in the virtual NL, SL and BS regimes, respectively. Allowing $q_1$ and $q_3$ to vary, we generate three virtual tumour populations of size $N=250$ by randomly selecting $N$ $(q_1,q_3)$ pairs, which correspond to the NL, SL and BS regimes, respectively. 

We then define the RT protocols of interest. We vary the dose amount $D \in \llbracket 0,5 \rrbracket \, (\si{Gy})$ and the number of doses per week $N_{frac} = \{ 1, 3,5\}$.  We assume that each dose is administered in $\delta_R = 10 \,\si{min}$ and, therefore, we vary the dimensionless dose rate, $R := \frac{D}{\delta_R R_{max}} \in \llbracket 0.1,0.5 \rrbracket$. We also suppose that all fractions are applied at the same time of day and the first weekly fraction is applied on Mondays, with subsequent fractions applied at equally spaced time intervals during Monday to Friday (e.g.  $3$ doses per week corresponds to doses on Monday, Wednesday and Friday).  Further, the duration of each fractionation schedule is determined so that the total dose administered is $80 \, \si{Gy}$ (or the closest multiple of $D$ to $80 \, \si{Gy}$). 

For each set of tumours and each RT protocol, we solve Equations \eqref{eq:2.7}-\eqref{eq:2.10} numerically for $t \in (0,t^*], \; t^* > 0$, using ODE45, a single step MATLAB built-in solver for non-stiff ODEs that is based on an explicit Runge-Kutta (4,5) formula, the Dormand-Prince pair \citep{dormand1980family}. For simplicity, we impose the initial conditions 
\begin{equation}
(T,T_S,T_R,c) = (T^*,0,0,c^*),
\label{eq:4.1.3}
\end{equation}
where $T^*$ and $c^*$ are the steady state tumour volume and oxygen concentration in the absence of treatment. All RT parameters are fixed at the default values listed in Table \ref{tab:3.1}.  

For each simulation, we record $\bar{T}, \, \bar{T_S}$ and $\bar{T_R}$, the mean undamaged, damaged and dead cell volumes in the last week of treatment. We also define the percent change in (mean) viable and total cell volumes between the start and the end of treatment as follows
\begin{equation}
\Delta_{\text{viable}} := 100 \times \frac{(\bar{T}+\bar{T_S}) - T_0 }{T_0} \quad \text{and} \quad \Delta_{\text{total}} := 100 \times \frac{\bar{\Sigma} - \Sigma_0 }{\Sigma_0},
\label{eq:4.1.1}
\end{equation}
where
\begin{equation*}
\Sigma_0 = T_0 + V_0, \quad \Sigma = T + T_S + T_R + V_0.
\end{equation*}
We also quantify the end-of-treatment tumour composition (relative to the total tumour volume at the start of treatment) as follows
\begin{equation}
\%T := 100 \times \frac{\bar{T}}{\Sigma_0}, \quad \%T_S := 100 \times \frac{\bar{T_S}}{\Sigma_0}, \quad \%T_R := 100 \times \frac{\bar{T_R}}{\Sigma_0} \quad \text{and} \quad \%V_0 := 100 \times \frac{V_0}{\Sigma_0}.
\label{eq:4.1.2}
\end{equation}
We note that the variables defined in \eqref{eq:4.1.2} can be used to describe $\Delta_{total} = (\%T + \%T_S + \%T_R + \%V_0) - 100$.  Finally, we record $\bar{c}$, the mean oxygen concentration in the last week of treatment, and the post-treatment steady state values of all the dependent variables.

\subsection{Characterising tumour response to fractionated RT}
In this section, we investigate the response of tumours in the NL, SL and BS virtual populations to fractionated RT. For each regime, we initially study tumour response to a conventional fractionation schedule consisting of $5 \times 2$ Gy fractions per week for $8$ weeks. In particular, we determine the average response and explore how certain values of $q_1$, $q_3$ and $V_0$ generate extremal behaviour.  We also study the impact of the dose and dosing frequency on tumour response.  We consider monostable regimes before looking at the bistable regime.

\subsubsection{Tumours in monostable regimes: the NL and SL virtual tumour populations}
\begin{figure}[!ht]
\begin{subfigure}[b]{0.5\textwidth}
\centering
\includegraphics[scale=0.4]{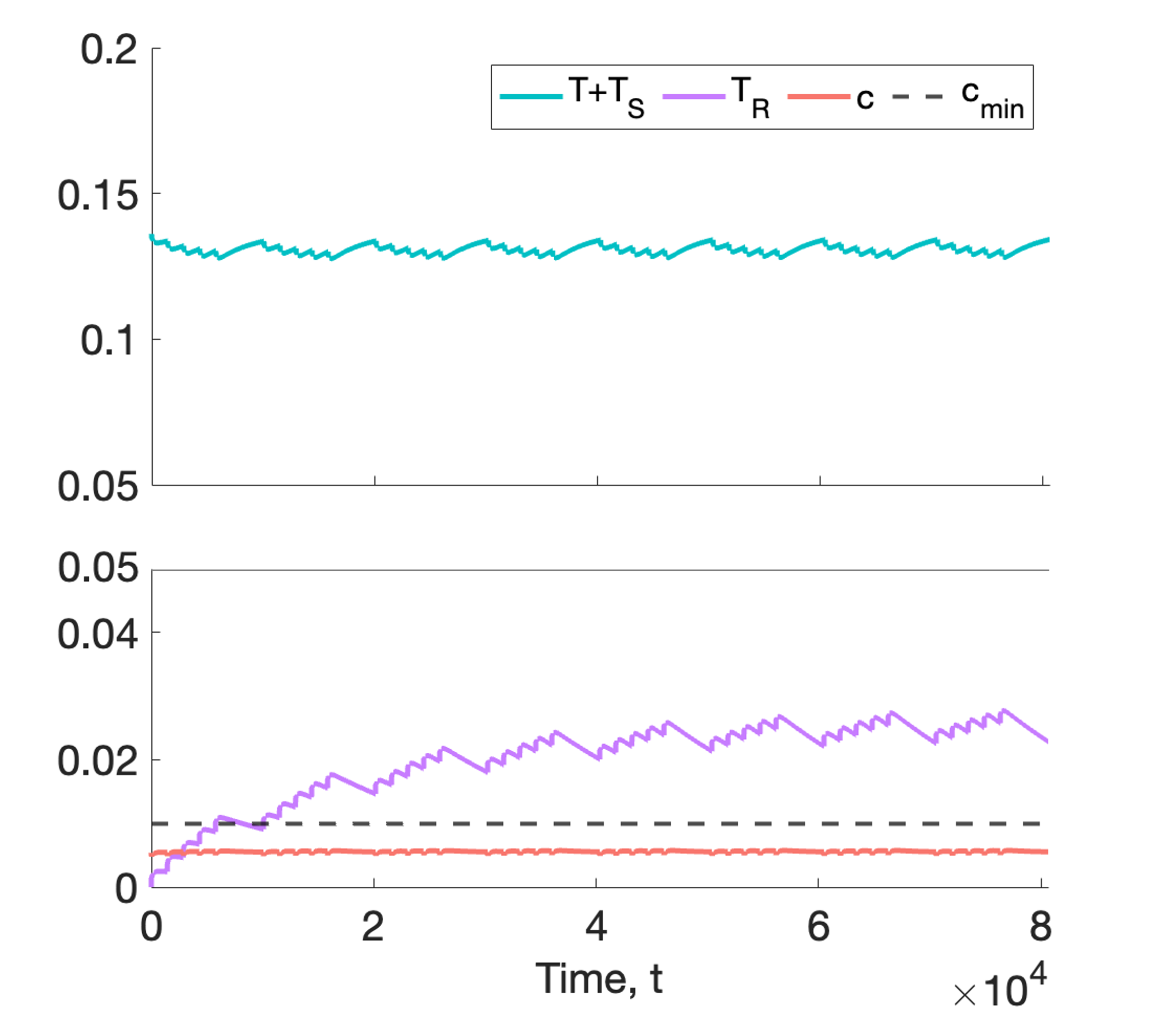} 
\subcaption{}
\label{fig:4.2.1A}
\end{subfigure}
\hfill
\begin{subfigure}[b]{0.5\textwidth}
\centering
\includegraphics[scale=0.4]{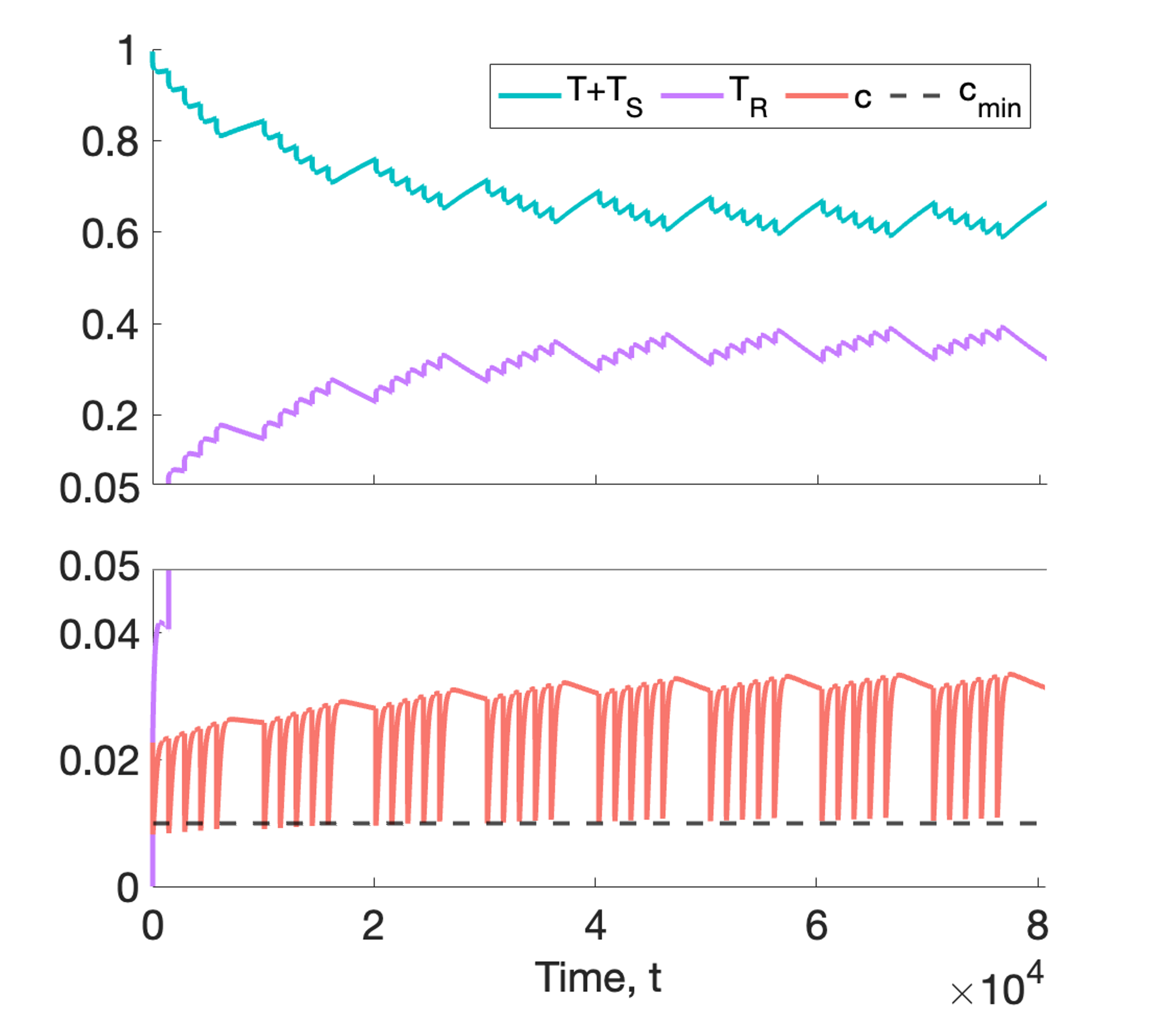} 
\subcaption{}
\label{fig:4.2.1B}
\end{subfigure}
\caption{For a conventional fractionation schedule, we numerically solve Equations \eqref{eq:2.7}-\eqref{eq:2.10} for $t \in (0,8 \times 10^4]$ subject to the initial conditions \eqref{eq:4.1.3}. In (a), we set $(q_1,q_3,V_0) = (0.832, 2.98,0.0005)$,  which corresponds to a NL tumour. In (b), we set $(q_1,q_3,V_0) = (1.08, 8.83,0.005)$,  which corresponds to a SL tumour. Although both tumours exhibit a decrease in viable cell volume, RT cell kill and accumulation of dead material is more significant for the SL than the NL tumour for this choice of parameter values.}
\label{fig:4.2.1}
\end{figure}

\paragraph{Typical responses to a conventional fractionation schedule.} Figure \ref{fig:4.2.1} shows the response of two NL and SL tumours to RT, the viable tumour cell volume, $T+ T_S$, of both decreasing during treatment.  Since, in both cases, the dependent variables evolve to time periodic solutions within $5$ weeks of treatment, we deduce that there is a maximal reduction in the viable cell volume that can be achieved with this fractionation schedule.  This maximum reduction, which we quantify using $\Delta_{viable}$, is significantly larger for the SL tumour at approximately $37.6\%$ than for the NL tumour at approximately $4.36\%$.  RT is more effective for the SL tumour as it is better oxygenated, and hence there is a higher rate of RT cell kill and greater accumulation of dead material, $T_R$.

Figure \ref{fig:4.2.1} also shows that, for both tumours, the oxygen concentration and the viable tumour cell volume decrease when RT is applied. This is because $T$ and $T_S$ cells consume oxygen at different rates: we recall that the oxygen consumption rates of sub-lethally damaged cells satisfy $q_{1,S} = 10 q_1$ and $q_{3,S} = 0.1 q_3$. Therefore, changes in tumour composition during treatment will alter the overall oxygen consumption rate of viable tumour cells, leading to transient, or persistent, increases or decreases in the oxygen concentration depending on the values of $q_1$ and $q_3$. 

\begin{figure}[!ht]
\centering
\includegraphics[scale=0.45]{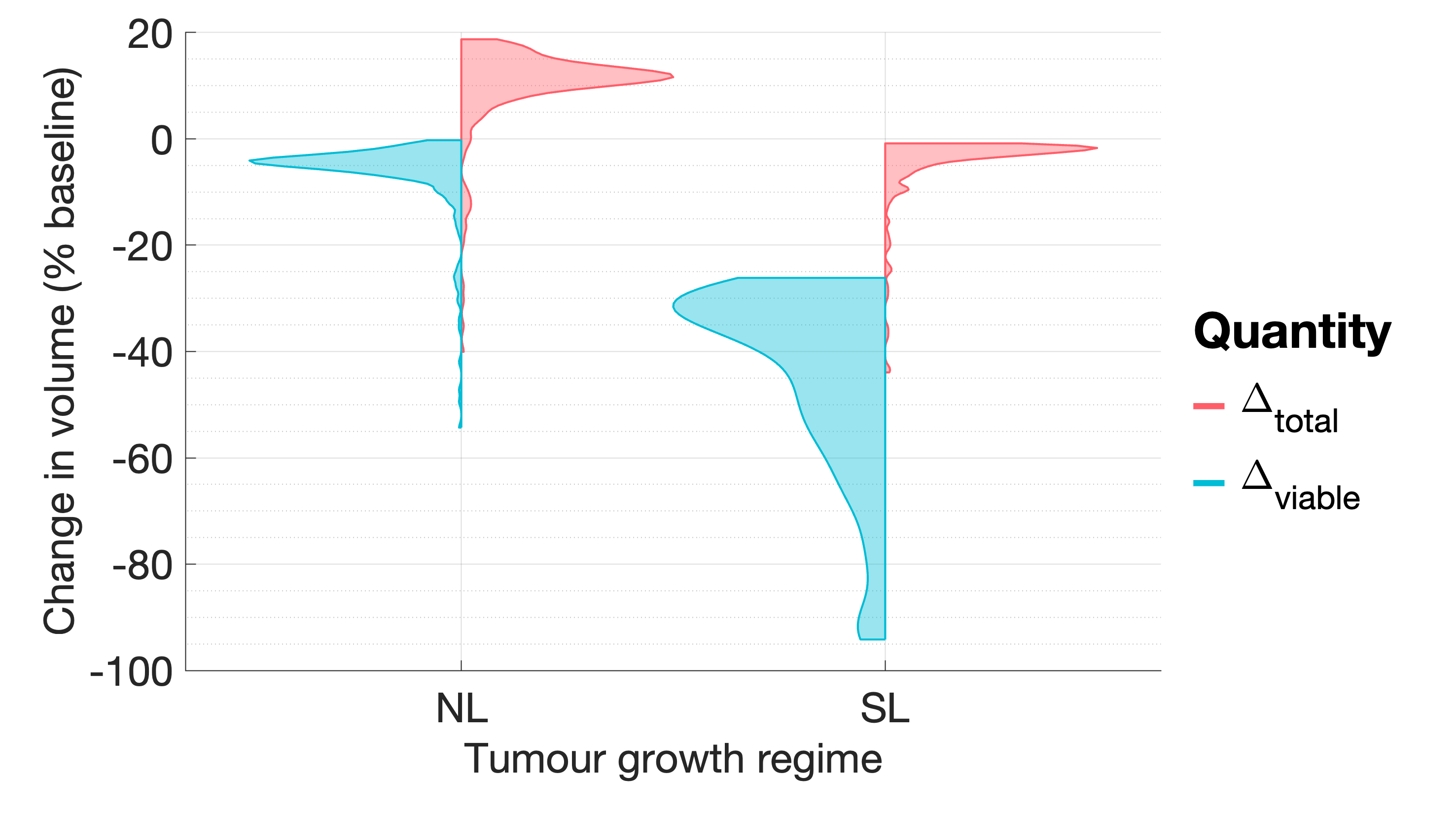} 
\caption{For virtual cohorts of NL and SL tumours, the violin plots show the distributions of $\Delta_{viable}$ and $\Delta_{total}$. The viable cell volume of all NL and SL tumours decreases during RT, with SL tumours showing significantly greater percentage changes. The total volume decreases for all SL tumours, while it increases for most NL tumours.We identify several outliers, which exhibit significantly larger reductions in their viable and total cell volumes. }
\label{fig:4.2.2}
\end{figure}

Figure \ref{fig:4.2.2} shows the distributions of $\Delta_{viable}$ and $\Delta_{total}$, following a conventional fractionation schedule, across the NL and SL virtual populations. We note that the behaviour shown in Figure \ref{fig:4.2.1} for specific NL and SL tumours is representative of the average behaviour of each virtual population.  In particular, tumours in the SL cohort typically respond well to treatment, with median and $(Q1,Q3)$ values of $\Delta_{viable}$ equal to $-37.9$ and $(-30.7,-54.1)$, respectively, and $\Delta_{total} < 0$ across the virtual population. Given the initial conditions \eqref{eq:4.1.3}, the latter follows because SL tumours fully occupy the available space at the start of treatment. Tumours in the NL cohort typically respond less well to treatment, with larger median and $(Q1,Q3)$ values of $\Delta_{viable}$ equal to $-4.57$ and $(-3.52,-6.53)$, respectively, and $\Delta_{total} > 0$, for at least $90\%$ of tumours.  When the net RT-induced cell death is minimal, NL tumours, which do not occupy all available free space at the start of treatment, can grow larger due increases in the dead cell volume. We also note that the value of $\Delta_{total}-\Delta_{viable}$ is larger for SL tumours since they accumulate more dead material.  

In both regimes, we observe outliers, which undergo much larger reductions in $T+T_S$ and $\Sigma$ than the average tumour. This suggests that certain parameter values within the NL and SL regimes correspond to tumours which are more sensitive to RT than the average NL and SL tumour.  

\begin{figure}[!h]
\begin{subfigure}[c]{0.5\textwidth}
\centering
\includegraphics[scale=0.45]{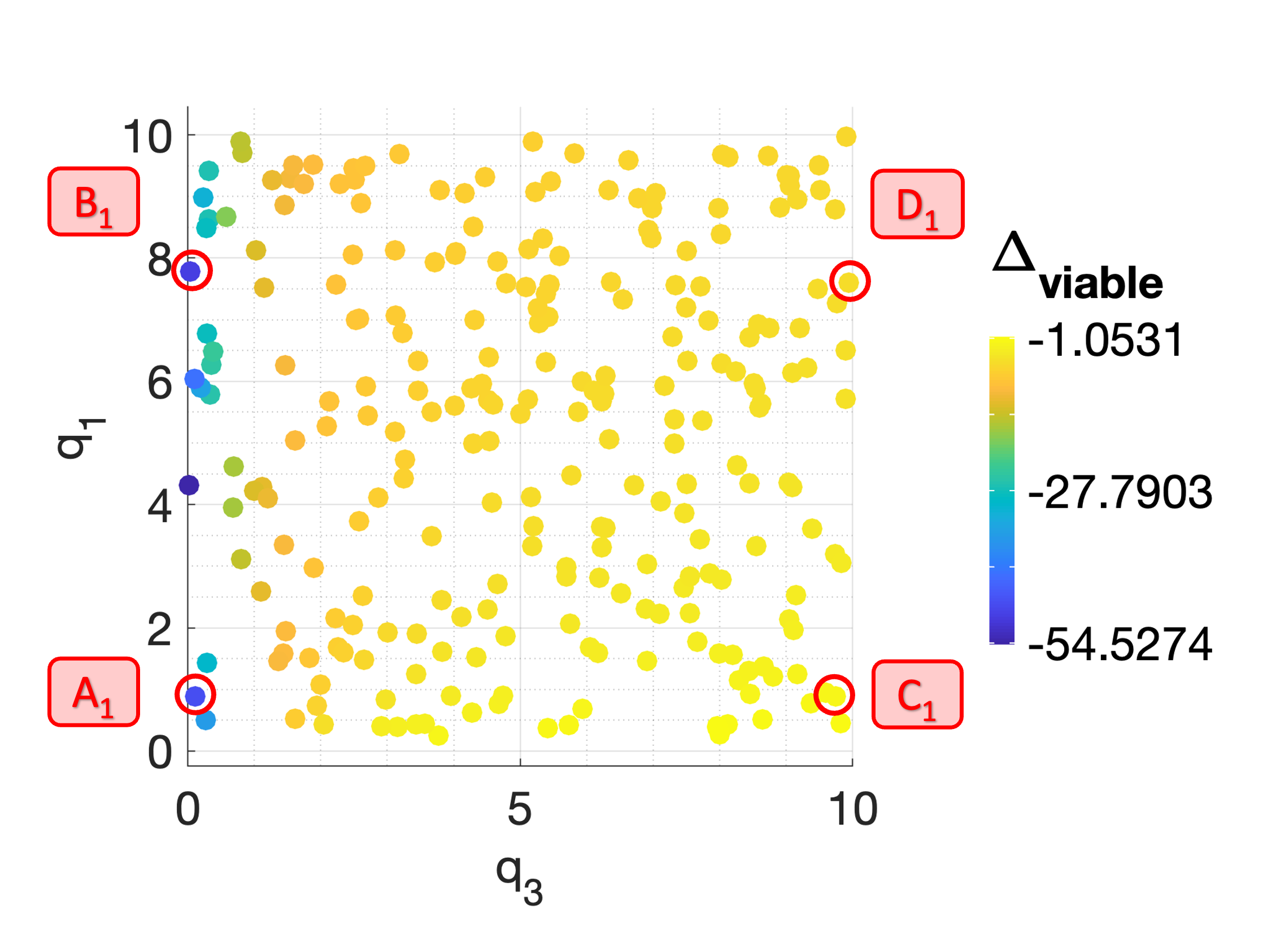} 
\label{fig:4.2.3A}
\end{subfigure}
\hfill
\begin{subfigure}[c]{0.5\textwidth}
\centering
\includegraphics[scale=0.45]{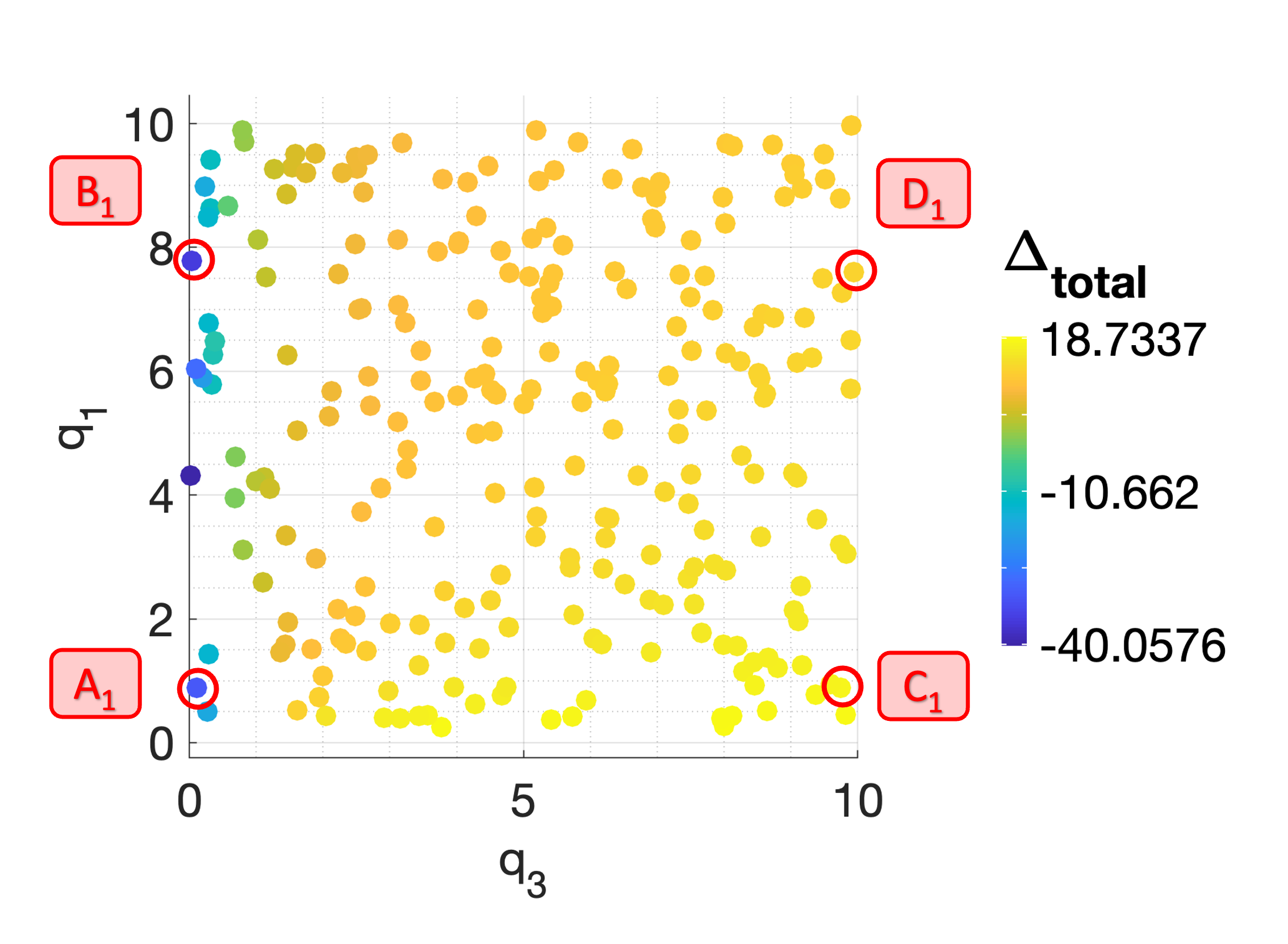} 
\label{fig:4.2.3B}
\end{subfigure}
\caption{The scatter plots show the values of $\Delta_{viable}$ and $\Delta_{total}$, following a conventional fractionation schedule, for each $(q_1,q_3)$ pair used to generate the set of virtual NL tumours.  $\Delta_{viable}$ and $\Delta_{total}$ increase with $q_3$ and decrease with $q_1$.}
\label{fig:4.2.3}
\end{figure}

\paragraph{The influence of the oxygen consumption rates, $q_1$ and $q_3$, on treatment outcome following a conventional fractionation schedule.} We now investigate the role of $q_1$ and $q_3$ in tumour response to RT.  The scatter plots in Figure \ref{fig:4.2.3} show the values of $\Delta_{viable}$ and $\Delta_{total}$ across the $(q_1,q_3)$ pairs which define the NL virtual population.  The response of NL tumours is most sensitive to the value of $q_3$, with smaller values leading to greater reductions in viable and total cell volumes. Further, higher values of $q_1$ are also associated with larger reductions in viable and total cell volumes. To understand these findings, we study the response to RT of four representative tumours corresponding to $(q_1,q_3)$ sets, $A_1$, $B_1$, $C_1$ and $D_1$ (see Figure \ref{fig:4.2.3} and Table \ref{tab:4.1}).

\begin{figure}[!h]
\centering
\includegraphics[scale=0.55]{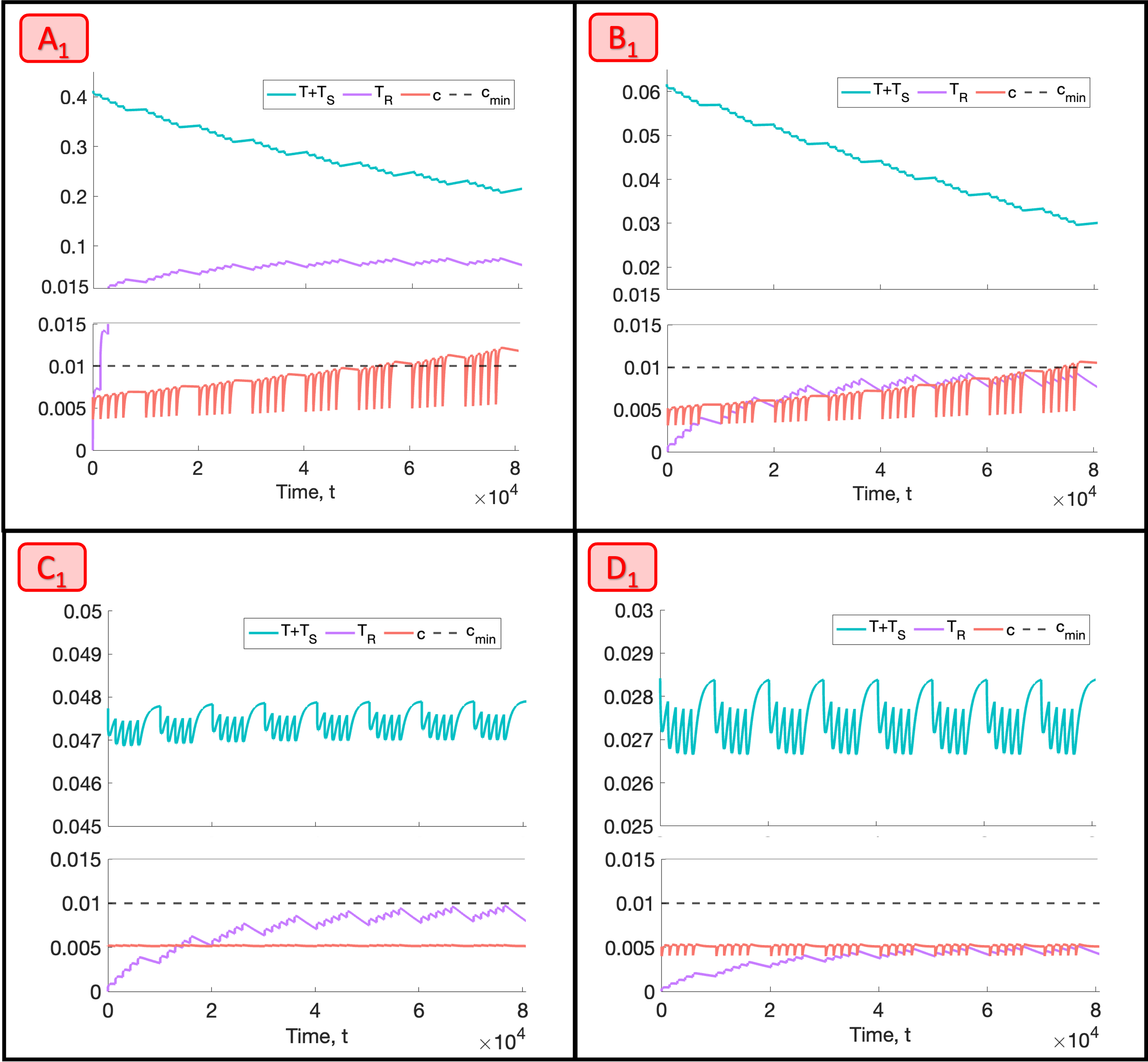} 
\caption{For a conventional fractionation schedule, we numerically solve Equations \eqref{eq:2.7}-\eqref{eq:2.10} for $t \in (0,8 \times 10^4]$ subject to the initial conditions \eqref{eq:4.1.3}. In $A_1$-$D_1$, we fix $V_0 = 0.0005$ and $(q_1,q_3)$ as indicated by the points $A_1$, $B_1$, $C_1$ and $D_1$ in Figure \ref{fig:4.2.3}, which correspond to NL tumours.  Comparing $A_1$-$D_1$ indicates that tumours with small values of $q_3$ ($A_1$, $B_1$) undergo a sustained decrease in $T+T_S$ during treatment whereas those with high values of $q_3$ ($C_1$, $D_1$) experience transient reductions in $T+T_S$ and significant regrowth between RT doses.}
\label{fig:4.2.4}
\end{figure}

\begin{figure}[!h]
\centering
\includegraphics[scale=0.45]{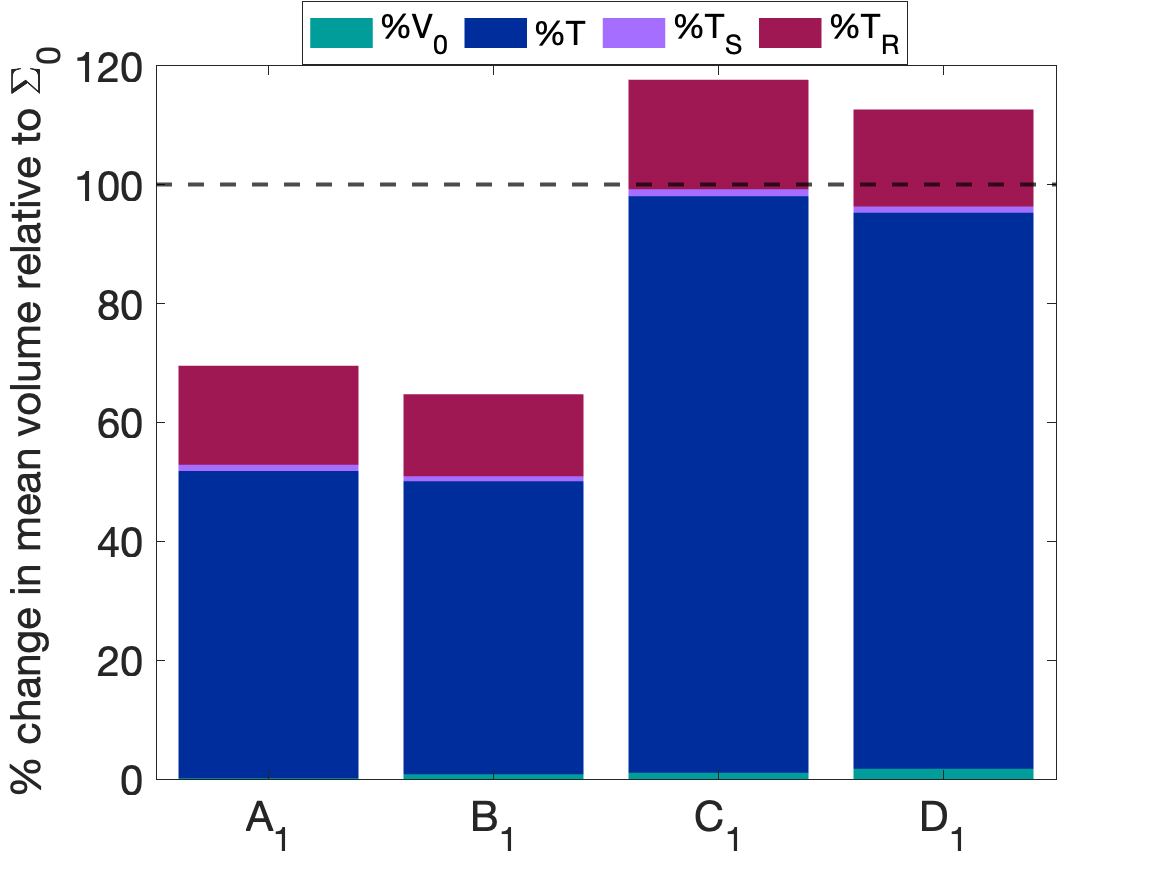} 
\caption{Bar graph showing the mean composition of tumours $A_1$-$D_1$ in the last week of a conventional fractionation schedule, where $\%T$, $\%T_S$, $\%T_R$ and $\%V_0$ are defined in \eqref{eq:4.1.2}. The tumours which undergo the largest decreases in viable cell and total volumes are characterised by a low value of $q_3$. A high value of $q_1$ also improves tumour response, but does not give rise to a large reduction in tumour volume. }
\label{fig:4.2.5}
\end{figure}

Comparing the response of tumours $A_1$ and $B_1$ to tumours $C_1$ and $D_1$ in Figure \ref{fig:4.2.4}, we see that a smaller value of $q_3$ implies higher average oxygen levels and slower cell proliferation (since $q_2 = 0.01 q_3$). We conclude two mechanisms could explain the increased efficacy of RT for low values of $q_3$: (i) higher rates of RT cell kill due to increased oxygenation or (ii) limited regrowth between RT fractions due to decreased proliferation.  

While oxygen levels are higher in tumours $A_1$ and $B_1$ than tumours $C_1$ and $D_1$ at the time of each dose of RT (see Figure \ref{fig:4.2.4}), their values of $\%T_R$ are slightly smaller (see Figure \ref{fig:4.2.5}). This suggests that the net increase in oxygen levels when values of $q_3$ are small does not significantly impact the proportion of cell kill due to RT.  By contrast, Figure \ref{fig:4.2.4} shows that the viable cell volume of tumours $A_1$ and $B_1$ increases marginally ($A_1$) or remains approximately constant ($B_1$) between fractions, whereas the viable cell volume of tumours $C_1$ and $D_1$ increases significantly between fractions, returning to its initial volume over the week-end break from RT.  This indicates that the value of $q_3$ impacts the reduction in the tumour burden by modulating tumour regrowth between fractions (rather than by increasing RT-induced cell death).

\begin{wraptable}{r}{0.5\textwidth}
\renewcommand{\arraystretch}{1.5}
\small
\begin{tabular}{c|c|c} \hline 
Tumour & $q_1$ & $q_3$ \\ \hline \hline
$A_1$ & $8.91 \times 10^{-1}$ & $1.14  \times 10^{-1}$  \\ 
$B_1$ &$ 7.78$ & $4. 01 \times 10^{-2}$ \\ 
$C_1$ & $8.91 \times 10^{-1}$ & $9.75$  \\ 
$D_1$ &  $7.60$ & $9.94$  \\ \hline
\end{tabular}
\caption{Parameter sets $A_1$, $B_1$, $C_1$ and $D_1$ corresponding to the representative NL tumours.}
\label{tab:4.1}
\end{wraptable}

Figure \ref{fig:4.2.5} also shows that a larger value of $q_1$ can slightly increase the magnitude of the reductions in $\Delta_{viable}$ and $\Delta_{total}$.  Since high values of $q_1$ lead to lower average oxygen levels (Figure \ref{fig:4.2.4}),  RT cell kill rates are smaller, while the rate of cell death due to hypoxia is larger than for low values of $q_1$.  The balance between these two processes determines whether cell death increases or decreases as $q_1$ increases. For tumours $C_1$ and $D_1$, Figure \ref{fig:4.2.4} shows that the reduction in $T+T_S$ following RT is greater and the increase in $T_R$ is smaller for larger values of $q_1$.  This confirms that a larger reduction in tumour burden can be achieved for large values of $q_1$ despite a reduction in RT-induced cell death: in such cases, increased cell death due to hypoxia drives the reduction in tumour volume.

Overall, we have shown that both low values of $q_3$ and high values of $q_1$ characterise the best NL responders. Since Figures \ref{fig:4.2.4} and \ref{fig:4.2.5} suggest that the value of $q_1$ has a less significant influence on tumour reduction than $q_3$, we conclude that growth limitation between RT fractions, rather than high rates of cell death due to RT or oxygen insufficiency, has the greatest influence on the efficacy of RT for NL tumours.

\begin{figure}[!h]
\begin{subfigure}[c]{0.5\textwidth}
\centering
\includegraphics[scale=0.45]{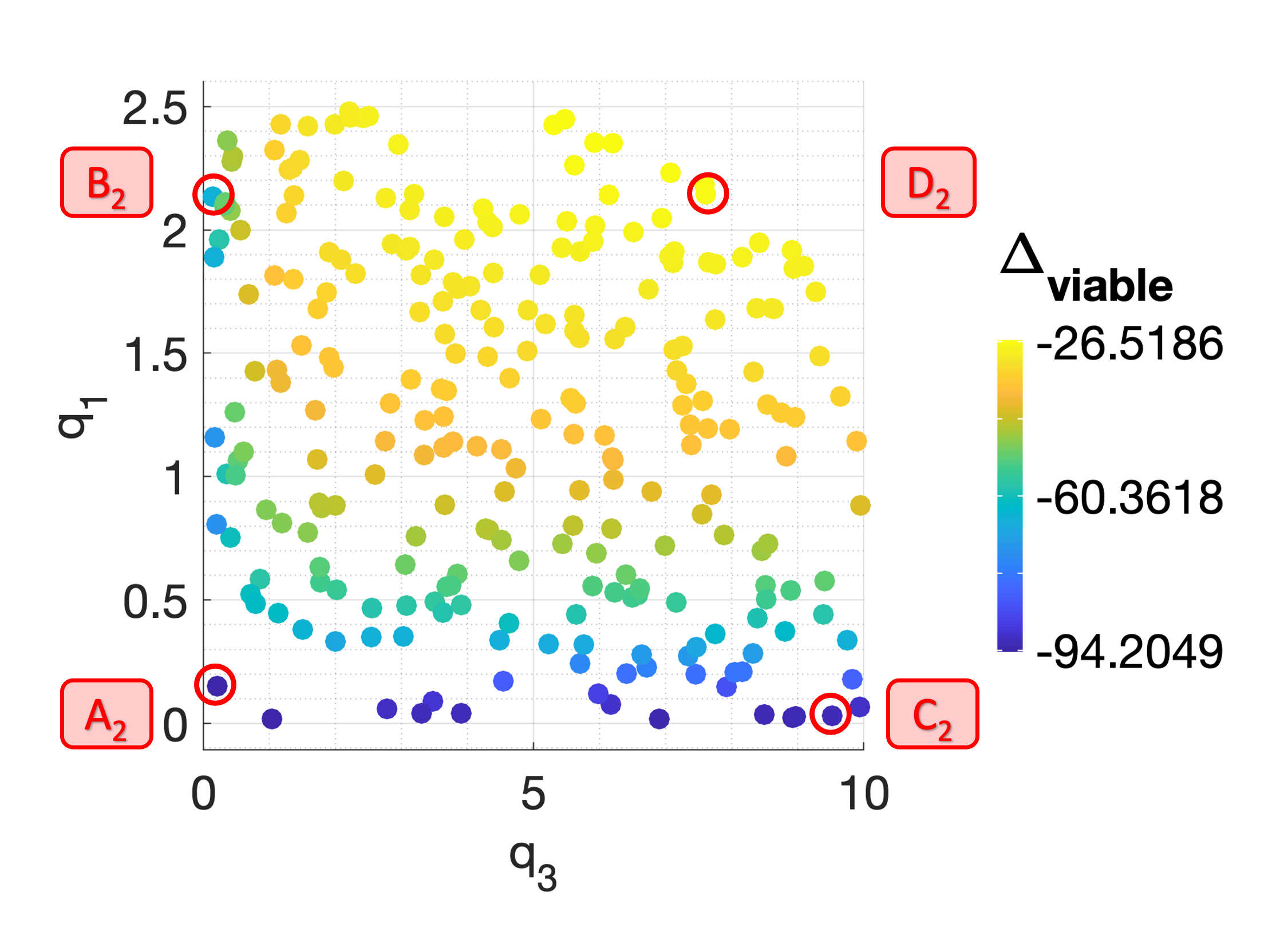} 
\label{fig:4.2.6A}
\end{subfigure}
\hfill
\begin{subfigure}[c]{0.5\textwidth}
\centering
\includegraphics[scale=0.45]{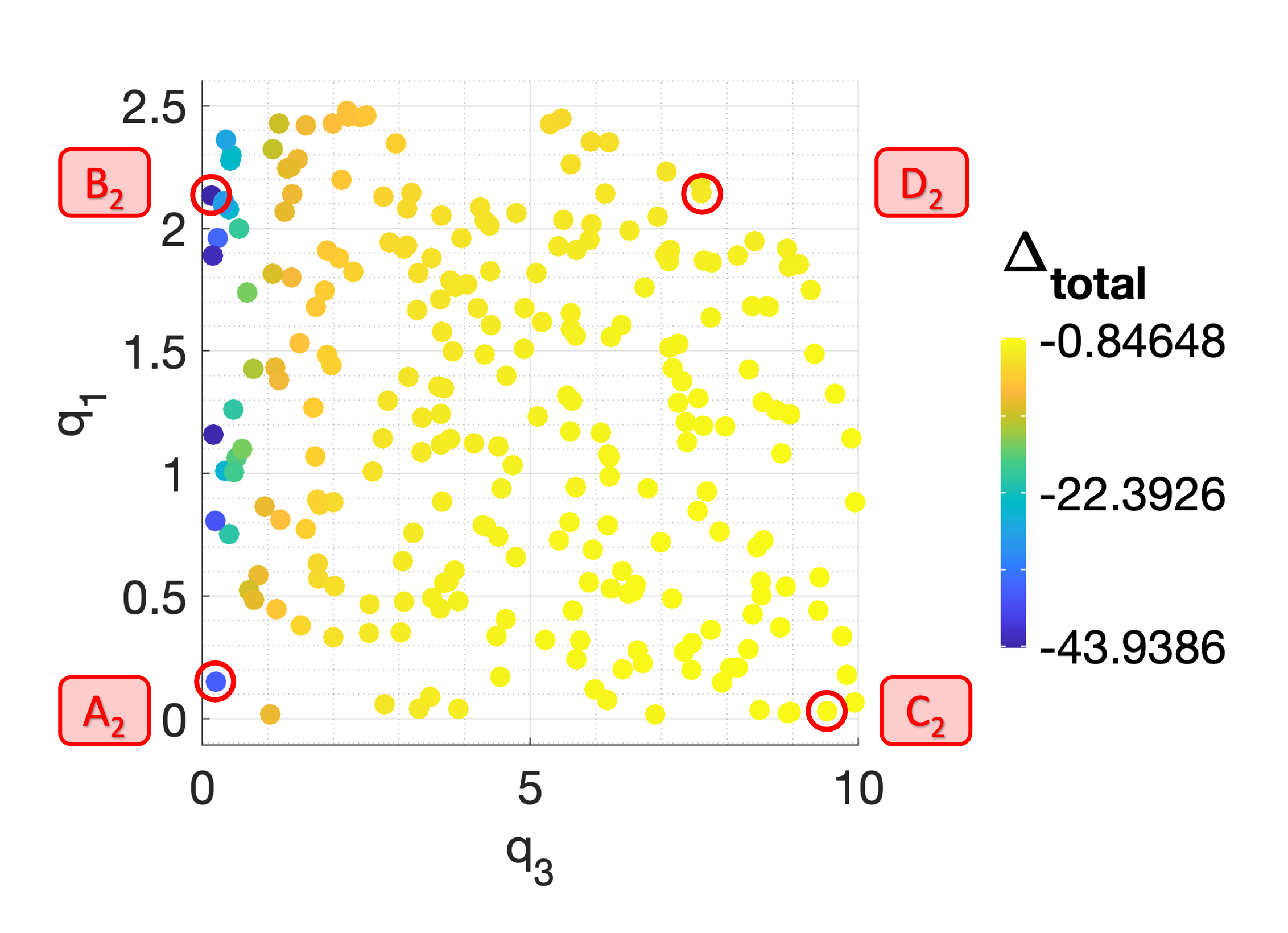} 
\label{fig:4.2.6B}
\end{subfigure}
\caption{The scatter plots show the values of $\Delta_{viable}$ and $\Delta_{total}$, following a conventional fractionation schedule, for the $(q_1,q_3)$ pairs used to generate the set of virtual SL tumours.  Smaller values of $\Delta_{viable}$ are obtained for low values of $q_1$ and/or $q_3$, while smaller values of $\Delta_{total}$ are obtained for low values of $q_3$.}
\label{fig:4.2.6}
\end{figure}

Figure \ref{fig:4.2.6} shows the values of $\Delta_{viable}$ and $\Delta_{total}$ across the $(q_1,q_3)$ pairs which define the SL virtual population.  The response of SL tumours is sensitive to the values of both $q_1$ and $q_3$: greater reductions in viable cell volume are obtained for smaller values of $q_1$ and/or $q_3$, while greater reductions in total cell volume are obtained for smaller values of $q_3$. To understand these results, we study the response to RT of tumours corresponding to four representative $(q_1,q_3)$ sets $A_2$, $B_2$, $C_2$ and $D_2$ (see Figure \ref{fig:4.2.6} and Table \ref{tab:4.2}).

\begin{figure}[!h]
\centering
\includegraphics[scale=0.45]{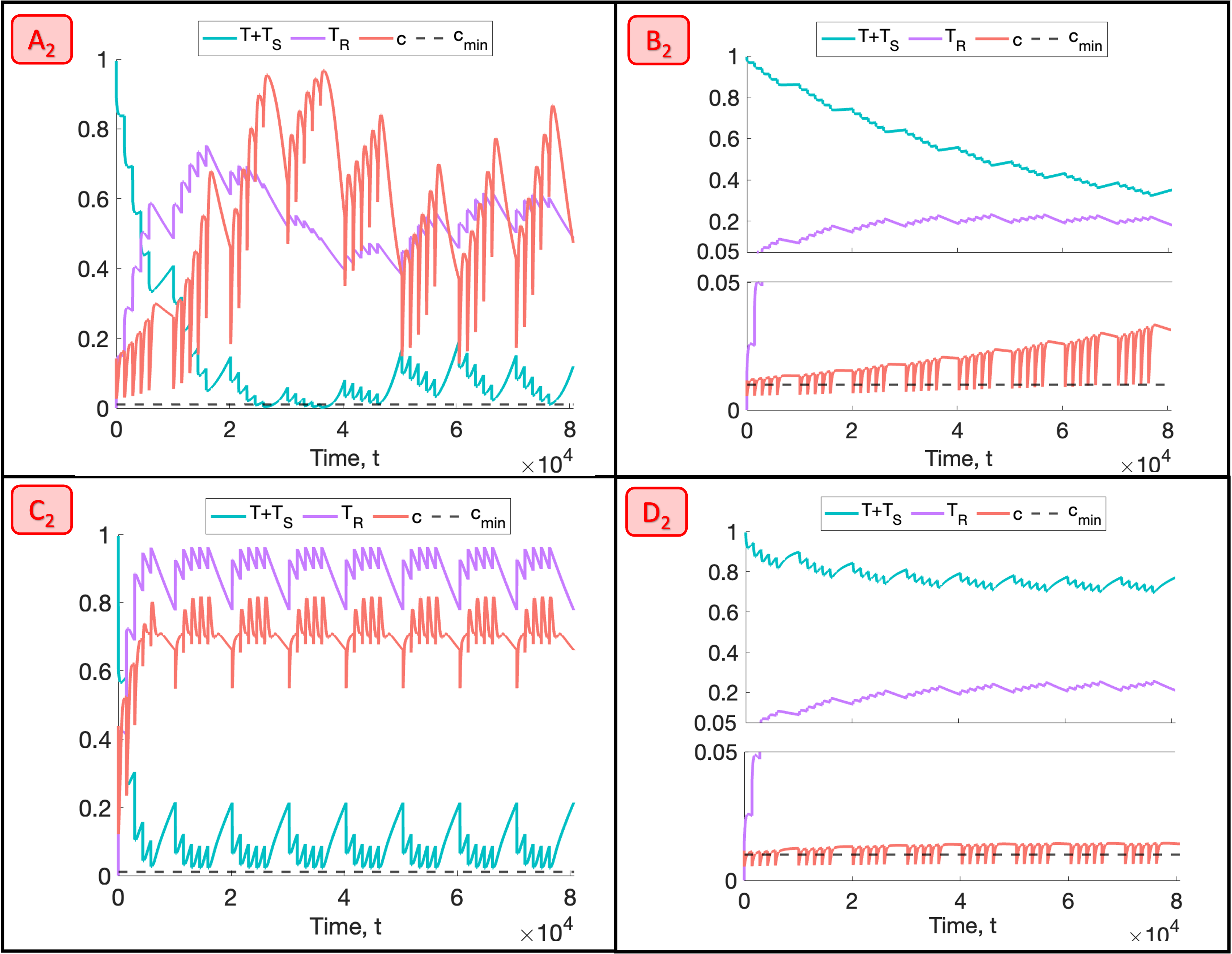} 
\caption{For a conventional fractionation schedule, we numerically solve Equations \eqref{eq:2.7}-\eqref{eq:2.10} for $t \in (0,8 \times 10^4]$ subject to the initial conditions \eqref{eq:4.1.3}. In $A_2$-$D_2$, we fix $V_0 = 0.005$ and $(q_1,q_3)$ as indicated by the points $A_2$, $B_2$, $C_2$ and $D_2$ in Figure \ref{fig:4.2.6}, which correspond to SL tumours. There are three qualitative behaviours: (i) low $q_1$ ($A_2$, $C_2$) is associated with high oxygen levels, and large RT cell kill and accumulation of dead cell material, (ii) a combination of low $q_3$ and high $q_1$ ($B_2$) implies limited inter-fraction tumour growth and (iii) a combination of high $q_1$ and $q_3$ ($D_2$) leads to low net cell death as proliferation rates are high and death rates are low.}
\label{fig:4.2.7}
\end{figure}

\begin{figure}[!h]
\centering
\includegraphics[scale=0.45]{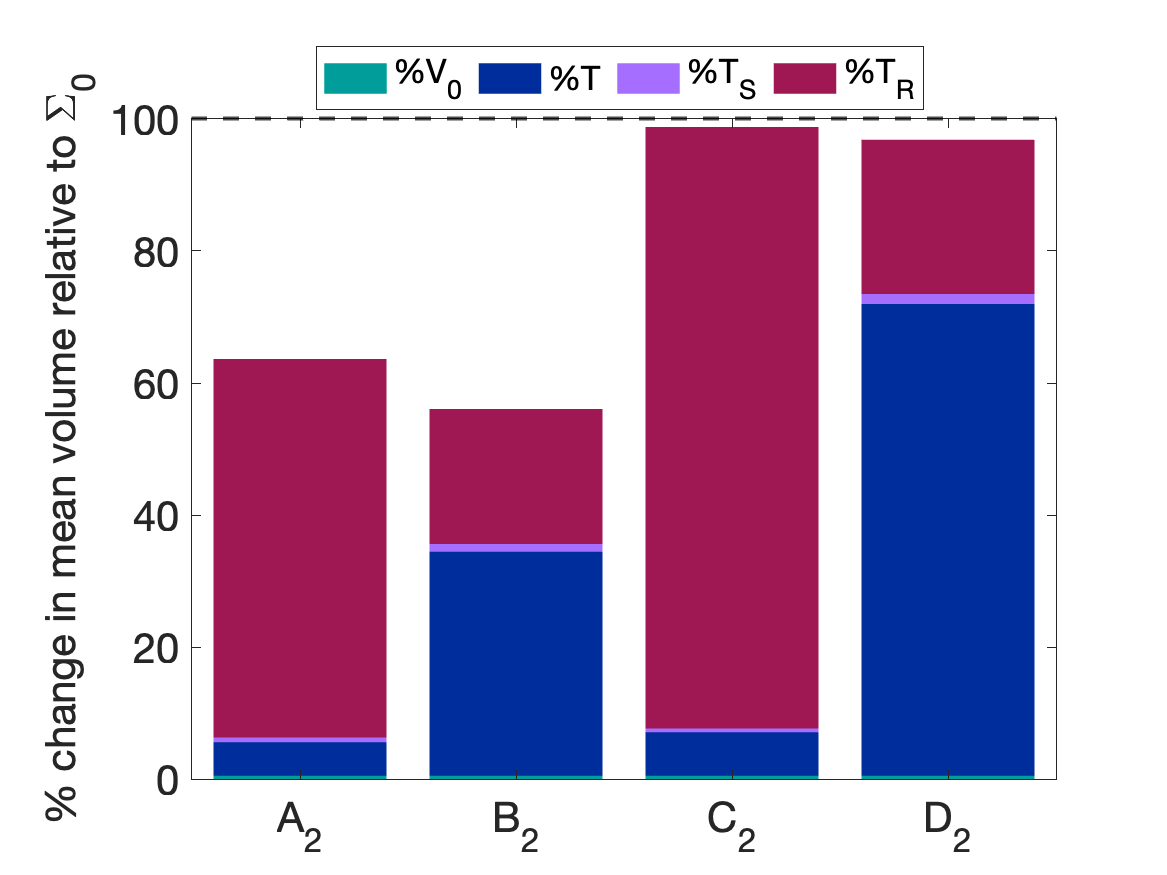} 
\caption{Bar graph showing the mean composition of tumours $A_2$-$D_2$ in the last week of a conventional fractionation schedule, where $\%T$, $\%T_S$, $\%T_R$ and $\%V_0$ are defined in \eqref{eq:4.1.2}. The value of $q_1$ influences tumour composition, whereas the value of $q_3$ determines the total tumour volume. In particular, for low values of $q_1$, $\%T_R$ is greater while $\%T_S$ is smaller, and, for low values of $q_3$, there is a greater reduction in $\Sigma$. }
\label{fig:4.2.8}
\end{figure}

Figures \ref{fig:4.2.7} and \ref{fig:4.2.8} reveal that $A_2$ and $C_2$ accumulate a larger number of dead cells than tumours $B_2$ and $D_2$.  This difference in tumour composition is amplified during treatment and the parameter which influences most this distinction is $q_1$. Figure \ref{fig:4.2.7} shows that for low values of $q_1$ (tumours $A_2$ and $C_2$), the intratumoural oxygen concentration, $c$, is at least 10-fold higher than for high values of $q_1$ (tumours $B_2$ and $D_2$). In particular,  $c \gg c_{min}$ throughout treatment when $q_1$ is small, which means that there is no cell death due to nutrient insufficiency and cell death is solely attributable to RT.  Therefore, the decrease in viable cell volume (and corresponding increase in dead cell volume) in tumours $A_2$ and $C_2$ following each RT fraction is driven by cell kill due to RT, which is enhanced by low values of $q_1$. 

\begin{wraptable}{r}{0.5\textwidth}
\renewcommand{\arraystretch}{1.5}
\small
\begin{tabular}{c|c|c}\hline
Tumour & $q_1$ & $q_3$ \\ \hline \hline
$A_2$ & $1.51 \times 10^{-1}$ & $2.10 \times 10^{-1} $  \\ 
$B_2$ &$ 2.14$ & $1.43 \times 10^{-1}$ \\ 
$C_2$ & $3.21 \times 10^{-2}$ & $9.53$  \\ 
$D_2$ &  $2.14$ & $7.61$  \\ \hline
\end{tabular}
\caption{Parameter sets $A_2$, $B_2$, $C_2$ and $D_2$ corresponding to the representative SL tumours.}
\label{tab:4.2}
\end{wraptable}

For tumours $B_2$ and $D_2$, Figure \ref{fig:4.2.7} also shows that, even though the oxygen concentration transiently drops below $c_{min}$ when each RT fraction is applied, there is a net increase in $c$ throughout treatment and, in particular, the weekly average oxygen concentration remains above $c_{min}$ (result not shown).  Therefore, we expect RT cell kill to increase during the fractionation schedule and cell death due to hypoxia to decrease. Since RT cell kill remains limited by low oxygen levels for both tumours, neither of the two proposed cell death mechanisms is responsible for the increased RT efficacy for tumour $B_2$ compared to tumour $D_2$.  However, Figure \ref{fig:4.2.7} reveals that $T+T_S$ increases at a slower rate between fractions for tumour $B_2$, which is characterised by low $q_3$. We, therefore, conclude that the increased RT efficacy is driven by reduced tumour regrowth between fractions (similarly to NL tumours with low $q_3$).  

Figure \ref{fig:4.2.8} further shows how low values of $q_3$ enable greater reductions in total tumour volume, $\Sigma$.  Since, for tumours $B_2$ and $D_2$, the values of $\%T_R$ are comparable while the value of $\%T$ is smaller for tumour $B_2$, the larger reduction in $\Sigma$ observed for tumour $B_2$ is due to increased net cell death (as described above).  In contrast,  for tumours $A_2$ and $C_2$, the values of $\%T$ are comparable while the value of $\%T_R$ is smaller for tumour $A_2$.  The larger reduction in $\Sigma$ observed for tumour $A_2$ is, therefore, due to a smaller accumulation of dead material, which occurs when lower viable cell volumes (caused by slower tumour regrowth between fractions) and/or lower oxygen levels reduce RT-induced cell death.

Overall, we have shown that two mechanisms can contribute to the increased efficacy of RT for certain tumours in a SL regime. These mechanisms are cell death due to RT and limited tumour regrowth between RT doses. Their relative contributions depend on the values of $q_1$ and $q_3$.  More specifically, when $q_1$ is small, RT cell kill is the dominant mechanism contributing to increased net cell death and, when $q_3$ is also small, limited regrowth between fractions ensures a larger reduction in total tumour volume.  When $q_1$ is large and $q_3$ is small, limited regrowth between fractions determines the response to RT by ensuring larger reductions in viable and total cell volumes. 

\begin{figure}[ht!]
\centering
\includegraphics[scale=0.4]{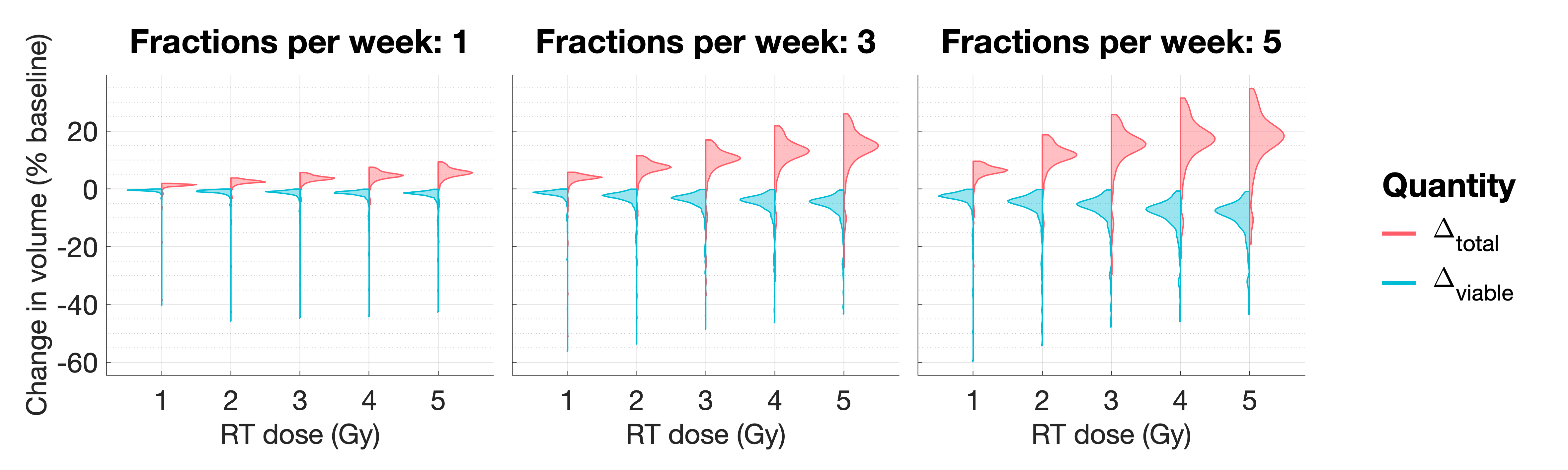} 
\caption{For the virtual NL population, we show how the distributions of $\Delta_{viable}$ and $\Delta_{total}$ change as the dose rate, $R \in \llbracket 0.1, 0.5 \rrbracket$, and the number of fraction per week, $N_{frac} \in \{1,3,5\}$, vary.  The mean value of $\Delta_{viable}$ decreases, while the mean values of $\Delta_{total}$ and $\Delta_{total} - \Delta_{viable}$ increase as $R$ and $N_{frac}$ increase.}
\label{fig:4.2.9}
\end{figure}

\begin{figure}[ht!]
\centering
\includegraphics[scale=0.4]{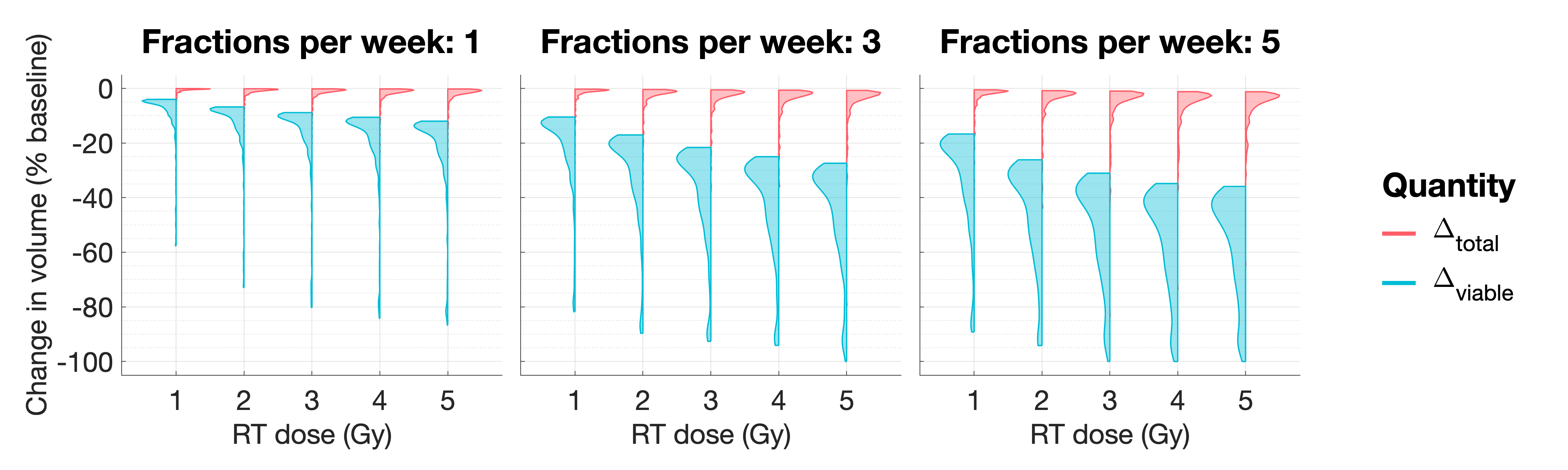} 
\caption{For the virtual SL population, we show how the distributions of $\Delta_{viable}$ and $\Delta_{total}$ change as the dose rate, $R \in \llbracket 0.1, 0.5 \rrbracket$, and the number of fraction per week, $N_{frac} \in \{1,3,5\}$, vary. The mean values of $\Delta_{viable}$ and $\Delta_{total}$ decrease, while the mean value of $\Delta_{total} - \Delta_{viable}$ increases as $R$ and $N_{frac}$ increase.}
\label{fig:4.2.10}
\end{figure}

\paragraph{The effect of the dosing schedule on typical tumour response.} We now consider how, for a fixed total dose, the dose rate, $R$, and the number of fractions per week, $N_{frac}$, affect tumour responses to RT.  For the virtual cohorts of NL and SL tumours, Figures \ref{fig:4.2.9} and \ref{fig:4.2.10}, respectively, show the distributions of $\Delta_{viable}$ and $\Delta_{total}$ for fractionation schedules with $R \in \llbracket 0.1,0.5 \rrbracket$ and $N_{frac} \in  \{ 1,3, 5\}$.  For SL tumours, we see that, on average, the reductions in the viable and total volumes and the difference between the viable and total volumes increase with $R$ and $N_{frac}$. The response of SL tumours is, thus, consistent with the current, standard approach to RT protocol design, which aims to maximise RT cell kill by applying a highest tolerable total dose, in sufficiently frequent fractions, to the tumour. This result is also supported by other modelling approaches, e.g., \citet{lewin2018evolution} developed a spatially resolved model of avascular tumour growth and RT cell death which predicted that there is a minimum RT dose, for a fixed dosing frequency, and a minimum dosing frequency, for a fixed RT dose, below which tumours grow during treatment. For NL tumours,  the mean reduction in viable volume and the difference between the viable and total volumes   also increase with $R$ and $N_{frac}$. However, the maximum reductions in viable and total volumes typically decrease with $R$ (for fixed $N_{frac}$), and the mean and maximum total volumes also increase with $R$ and $N_{frac}$.  Therefore, a higher dosing frequency and/or dose may not lead to greater RT efficacy.

\subsubsection{Tumours in the bistable regime} \label{sec:4.2.2}

\begin{figure}[!h]
\begin{subfigure}[t]{0.5\textwidth}
\centering
\includegraphics[scale=0.45]{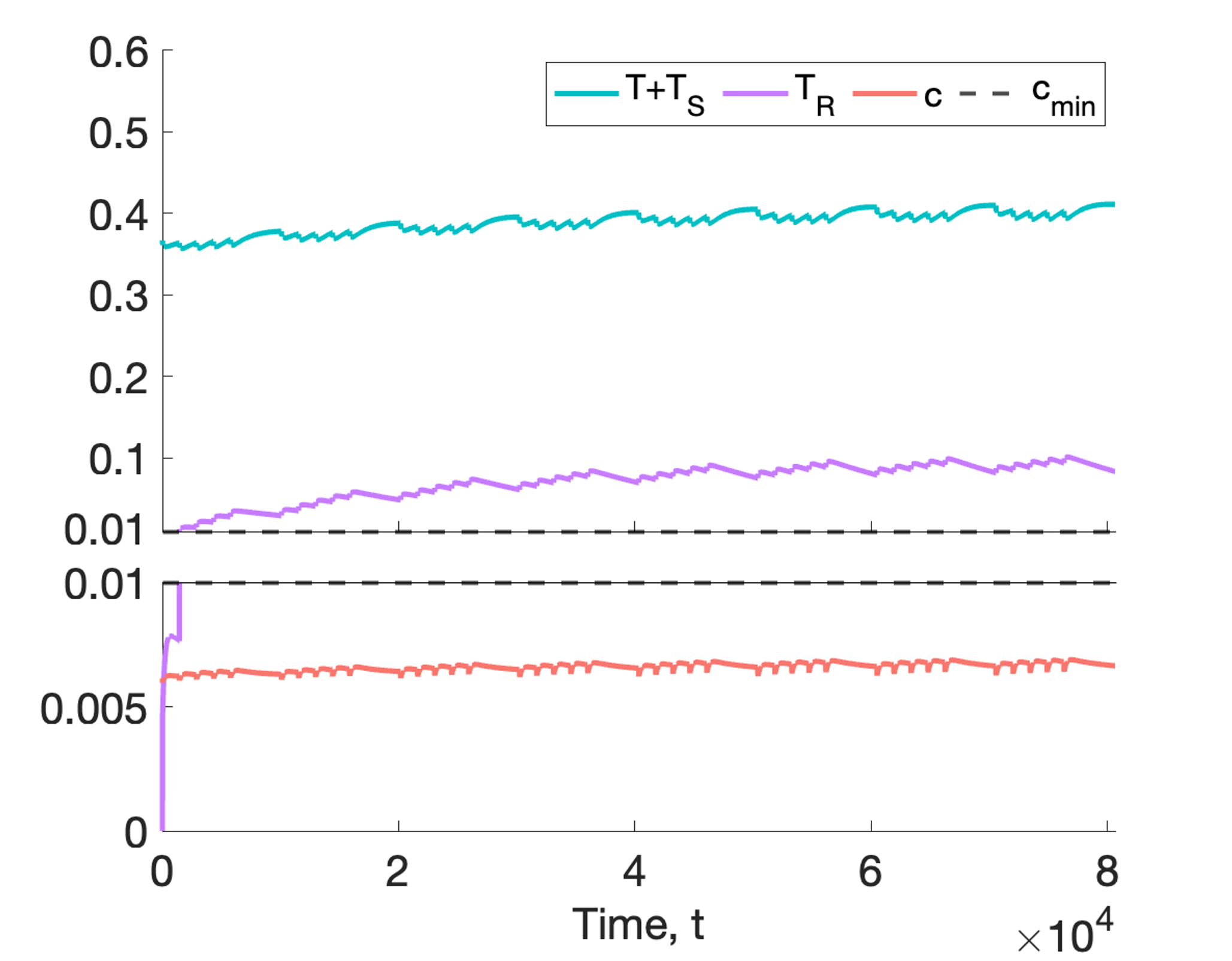} 
\caption{}
\label{fig:4.2.17A}
\end{subfigure}
\hfill
\begin{subfigure}[t]{0.5\textwidth}
\centering
\includegraphics[scale=0.4]{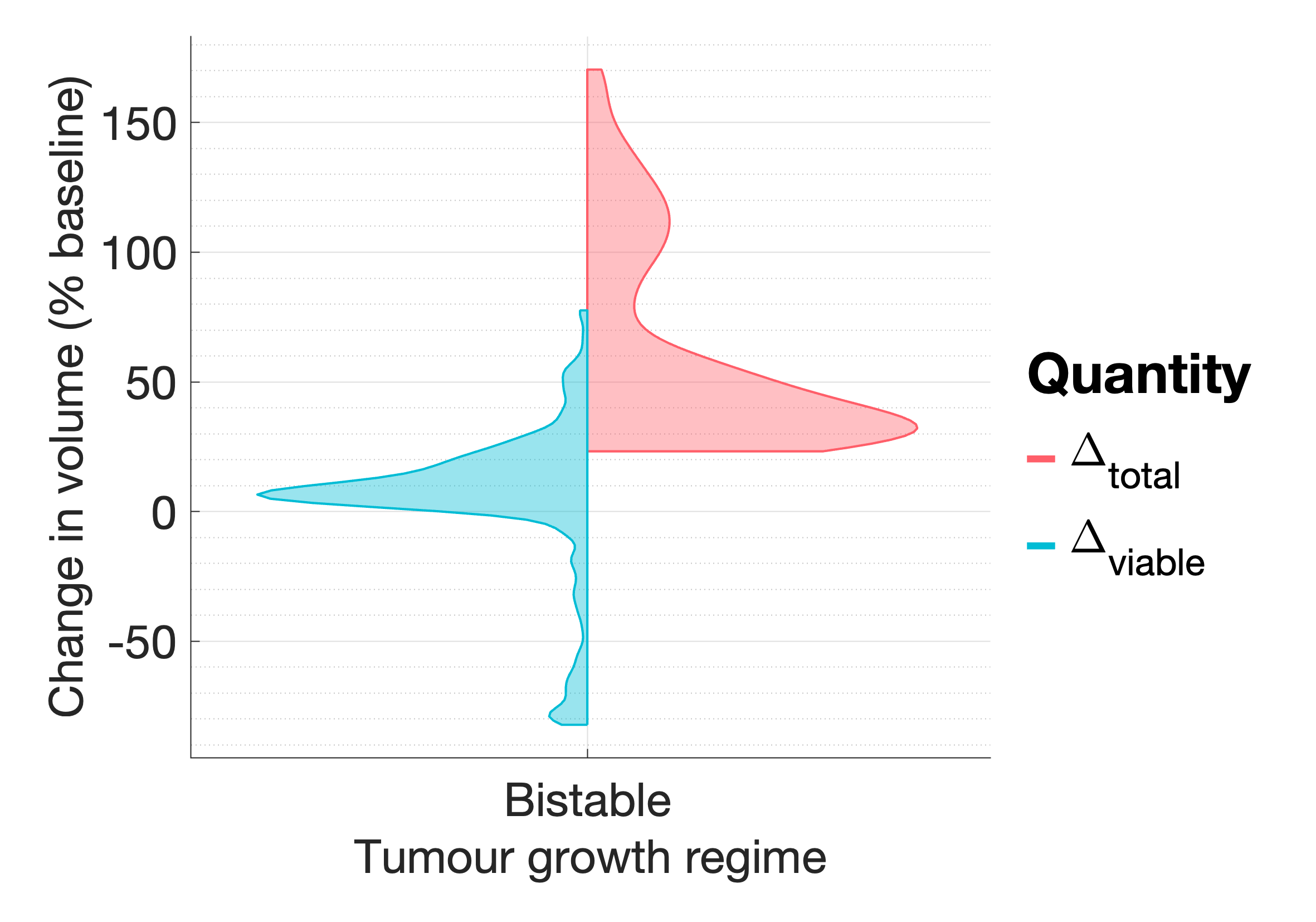} 
\caption{}
\label{fig:4.2.17B}
\end{subfigure}
\caption{(a) For a conventional fractionation schedule, we numerically solve the Equations \eqref{eq:2.7}-\eqref{eq:2.10} for $t \in (0,8 \times 10^4]$ subject to the initial conditions \eqref{eq:4.1.3}.  We set $(q_1,q_3,V_0) = (0.787, 8.38,0.00275)$. This tumour represents the typical behaviour in a BS regime.  (b) Violin plots representing the distributions of $\Delta_{viable}$ and $\Delta_{total}$. While the effect of RT is deleterious for most tumours, with several outliers experiencing larger than average increases in viable and total volumes, there are tumours that exhibit larger than average decreases in viable volume.}
\label{fig:4.2.17}
\end{figure}

\paragraph{Typical response to a conventional fractionation schedule.} Figure \ref{fig:4.2.17A} shows the average response of a tumour in a BS regime to a conventional fractionation schedule. RT has a detrimental effect as tumour regrowth between fractions and over the week-end outweighs RT-induced cell death. The dead cell volume also increases throughout treatment, implying an increase in total volume.  Figure \ref{fig:4.2.17B} further shows that, for the BS virtual cohort, $\Delta_{viable} > 0$ for at least $80\%$ of tumours and $\Delta_{total} > 0$ for all tumours. This reveals that most tumours in the BS virtual cohort respond badly to RT.

The results in Figure \ref{fig:4.2.17B} also indicate that a few virtual tumours are more or less sensitive to RT than the average tumour in the BS virtual population: while their total volume increases during RT, their viable volume undergoes a $20-80 \%$ decrease or $40-80\%$ increase, respectively, by the end of treatment. We investigate the response to RT of these outliers in more detail in the following section.

\begin{figure}[!h]
\centering
\includegraphics[scale=0.35]{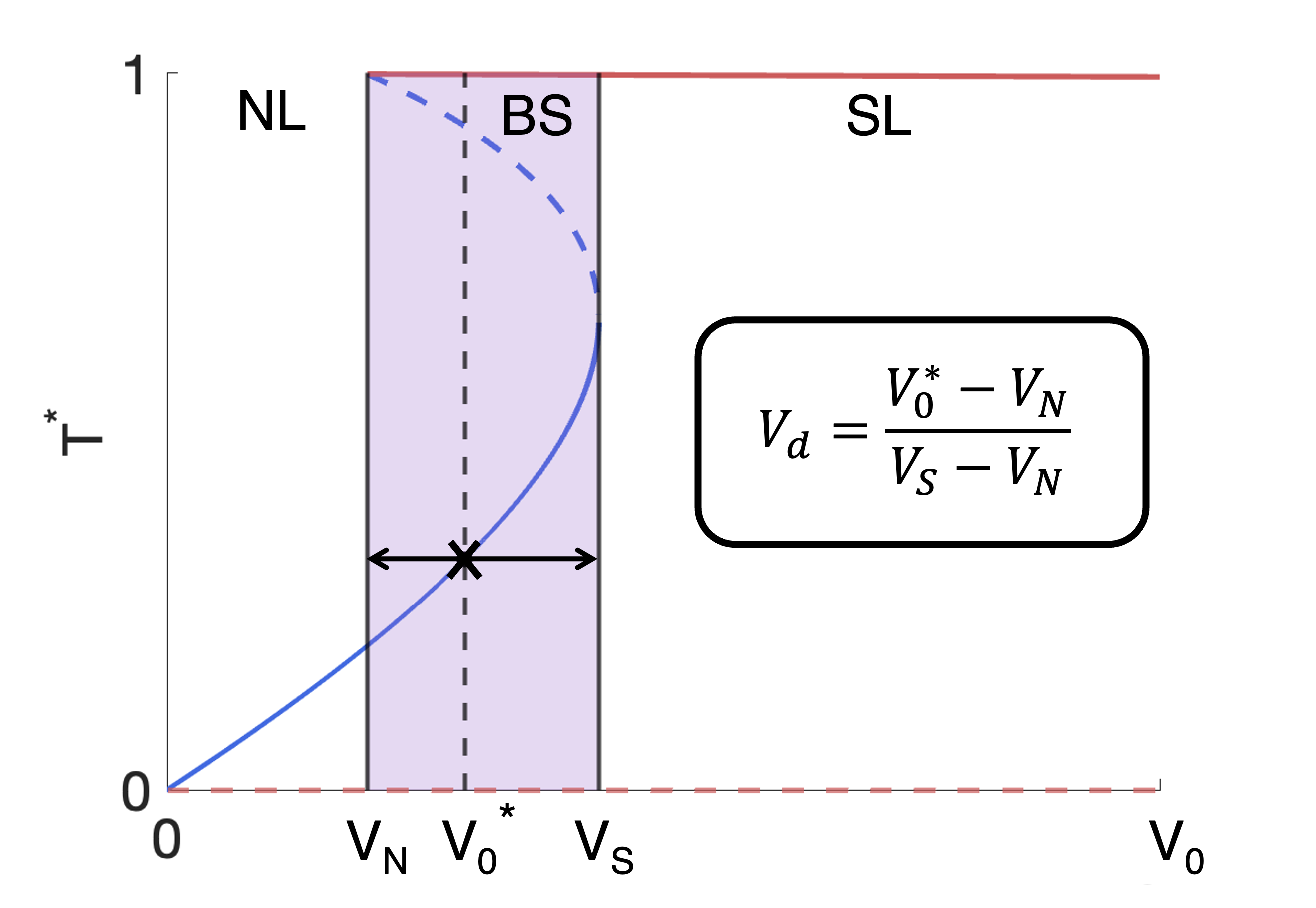} 
\caption{Schematic bifurcation diagram showing how, for fixed values of $q_1$ and $q_3$, the steady state value of the tumour cell volume, $T^*$,  changes with $V_0$.  The shaded purple region represents the bistable region, where $V_N$ and $V_S$ are the threshold values of $V_0$ below and above which only NL and SL steady states exist. For tumours with $V_N < V^*_0 < V_S$, we define $V_d$ by \eqref{eq:4.2.1} to quantify the relative proximity of $T^*(V^*_0)$ to the monostable NL and SL regions. }
\label{fig:4.2.14}
\end{figure}

\paragraph{The influence of $q_1$, $q_3$ and $V_0$ on treatment outcome following a conventional fractionation schedule.} As for tumours in monostable regimes, we study the influence of $q_1$ and $q_3$ on tumour response to RT, but we also study the role played by the vascular volume, $V_0$. More specifically, we introduce a function $V_d$, which quantifies how close a tumour in the BS regime lies to the NL and SL regimes (see the schematic in Figure \ref{fig:4.2.14}):
\begin{equation}
V_d(V_0) = \frac{V_0 - V_N}{V_S - V_N},
\label{eq:4.2.1}
\end{equation}
where $V_N$ and $V_S$ are the threshold values of $V_0$ below and above which only NL and SL steady states exist. Further,
\begin{equation}
\begin{cases}
V_d \to 0 \text{ as } V_0 \to V_N, \\
V_d \to 1 \text{ as } V_0 \to V_S.
\end{cases}
\label{eq:4.2.2}
\end{equation}

\noindent In particular, $V_d \gtrapprox 0$ for tumours which are close to the NL regime, whereas $V_d \lessapprox 1$ for tumours which are close to the SL regime.

\begin{figure}[!h]
\centering
\includegraphics[scale=0.4]{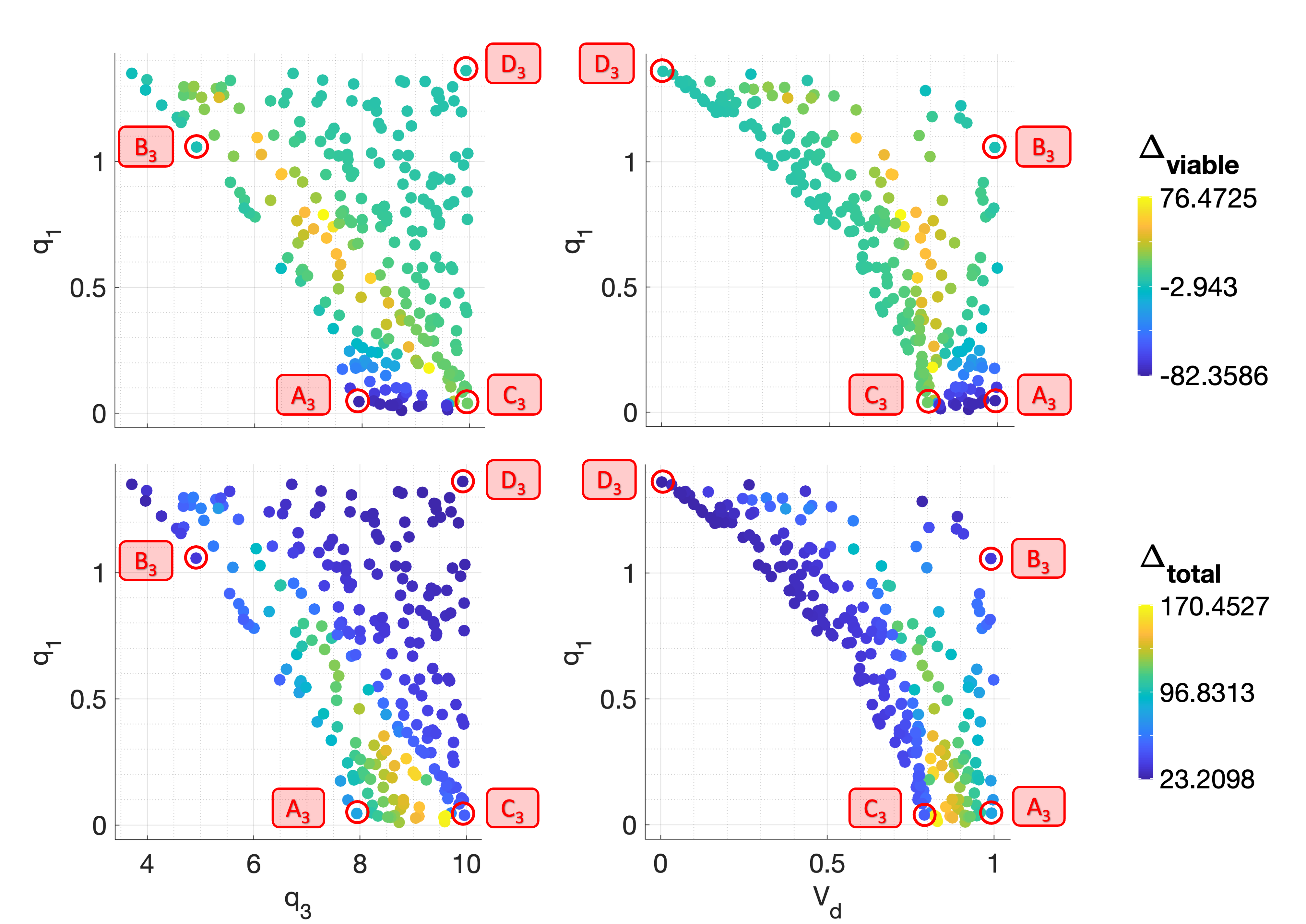} 
\caption{The scatter plots show the values of $\Delta_{viable}$ and $\Delta_{total}$, following a conventional fractionation schedule, for the $(q_1,q_3)$ and $(q_1,V_d)$ pairs used to generate the set of virtual tumours in a BS regime.  The smallest values of $\Delta_{viable}$ are associated with lower values of $q_1$ and $q_3$ and $V_d \lesssim 1$, while the largest values of $\Delta_{viable}$ are associated with higher values of $q_1$, lower values of $q_3$ and intermediate values of $V_d$. $\Delta_{total}$ is largest for the tumours with the smallest and largest values of $\Delta_{viable}$.}
\label{fig:4.2.11}
\end{figure}

The scatter plots in Figure \ref{fig:4.2.11} show the values of $\Delta_{viable}$ and $\Delta_{total}$ across the $(q_1,q_3)$ and $(q_1,V_d)$ pairs corresponding to the BS virtual population. We note that the values of $q_3$ and $V_d$ are correlated: for fixed $q_1$, the lowest value of $V_d$ corresponds to the highest values of $q_3$ and vice versa. It is, therefore, sufficient to describe the response of tumours in a BS regime with respect to the values $q_1$ and $V_d$. The largest reductions in viable volume are obtained for lower values of $q_1$ and $V_d \lesssim 1$, whereas the largest increases in viable volume are obtained for higher values of $q_1$ and intermediate values of $V_d$. Those tumours with the smallest and largest values of $\Delta_{viable}$ also undergo the largest increases in total volume: for intermediate to high values of $V_d$, $\Delta_{total}$ decreases as $V_d$ and $q_1$ increase.

\begin{figure}[!h]
\centering
\includegraphics[scale=0.45]{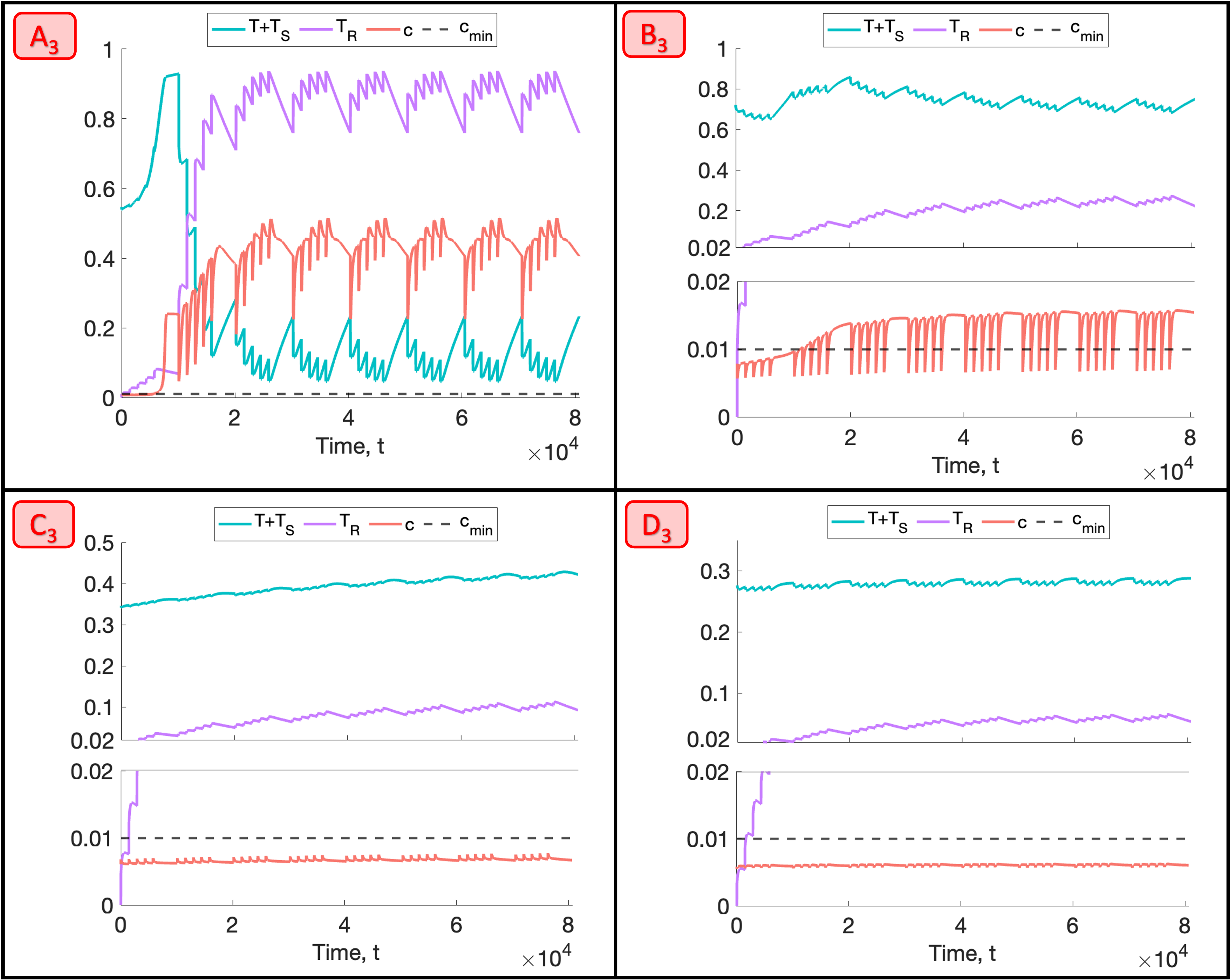} 
\caption{For a conventional fractionation schedule, we numerically solve Equations \eqref{eq:2.7}-\eqref{eq:2.10} for $t \in (0,8 \times 10^4]$ subject to the initial conditions \eqref{eq:4.1.3}. In $A_3$-$D_3$, we fix $V_0 = 0.00275$ and $(q_1,q_3)$ as indicated by the points $A_3$, $B_3$, $C_3$ and $D_3$ in Figure \ref{fig:4.2.11}.  $A_3$ and $B_3$ decrease in viable volume and increase in total volume, while $C_3$ and $D_3$ increase in both viable and total volumes. $A_3$ and $C_3$ are outliers. }
\label{fig:4.2.12}
\end{figure}

\begin{figure}[!h]
\centering
\includegraphics[scale=0.45]{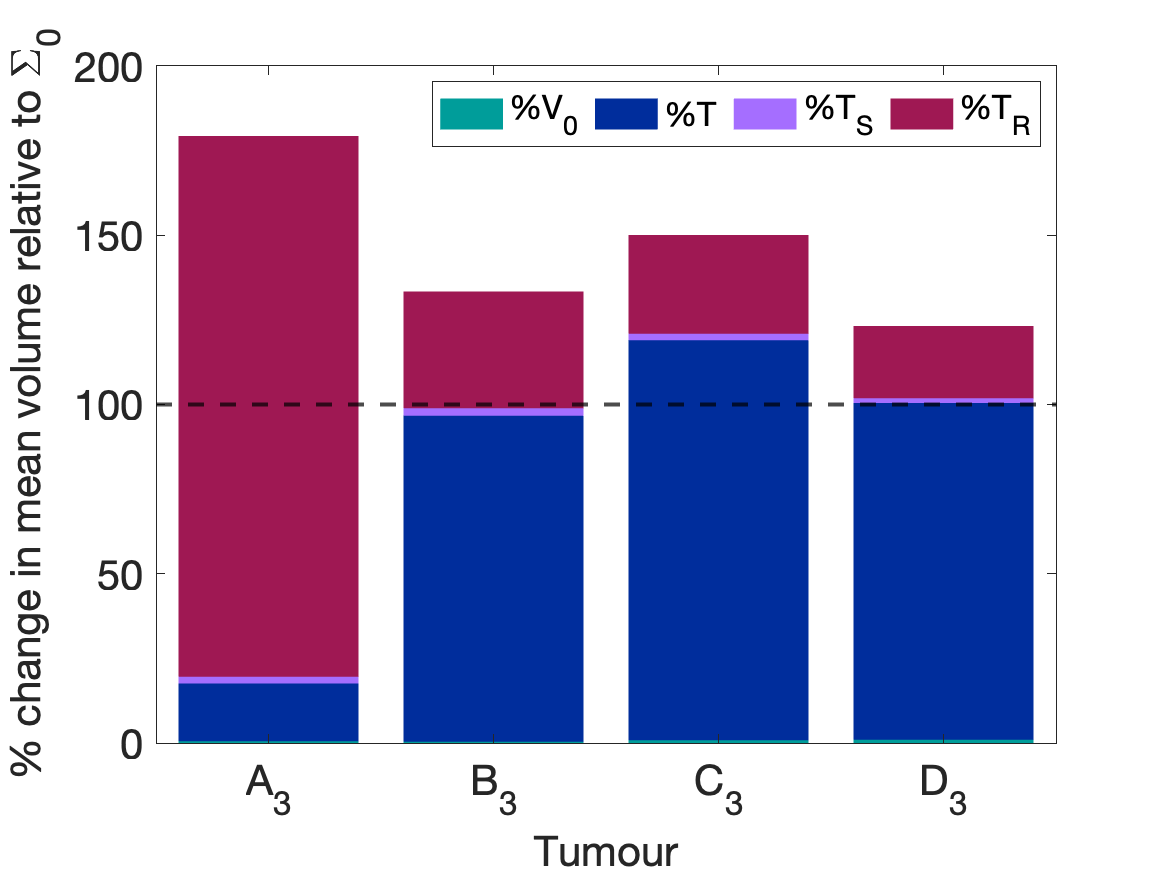} 
\caption{Bar graph showing the mean composition of tumours $A_3$-$D_3$ in the last week of a conventional fractionation schedule, where $\%T$, $\%T_S$, $\%T_R$ and $\%V_0$ are defined in \eqref{eq:4.1.2}.  A low value of $q_1$ and $V_d \approx 1$ ($A_3$) is necessary for larger than average decreases and increases in viable and total volumes, respectively. Larger than average increases in viable and total volumes are observed for intermediate values of $V_d$ ($C_3$).}
\label{fig:4.2.13}
\end{figure}

We now select four representative $(q_1,q_3, V_d)$ sets $A_3$, $B_3$, $C_3$ and $D_3$ (see Figure \ref{fig:4.2.11} and Table \ref{tab:4.3}) and study the corresponding tumours' responses to RT.  Figures \ref{fig:4.2.12} and \ref{fig:4.2.13} show that tumours $A_3$ and $B_3$ decrease in viable volume, with $A_3$ experiencing a larger than average reduction, while $C_3$ and $D_3$ increase in viable volume, with $C_3$ experiencing a larger than average increase. While tumours $A_3$ and $C_3$ have low $q_1$, $V_d \approx 1$ for $A_3$ and $V_d \approx 0.8$ for $C_3$.  Similarly, while tumours $B_3$ and $D_3$ have high $q_1$, $V_d \approx 1$ for $B_3$ and $V_d \approx 0$ for $D_3$. Given \eqref{eq:4.2.2}, this suggests that the behaviour of tumours in a BS regime that lie sufficiently close to the SL or NL regions will be, respectively, similar to that of SL or NL tumours with values of $q_1$ and $q_3$ of the same order of magnitude.  

\begin{wraptable}{r}{0.4\textwidth}
\renewcommand{\arraystretch}{1.5}
\small
\begin{tabular}{c|c|c|c}\hline
Tumour & $q_1$ & $q_3$ & $V_d$ \\ \hline \hline
$A_3$ & $4.55 \times 10^{-2}$ & $7.94 $ & $0.993 $ \\  
$B_3$ &$ 1.06$ & $4.92 $  & $0.991 $ \\  
$C_3$ & $3.92 \times 10^{-2}$ & $9.96$ & $0.791 $  \\  
$D_3$ &  $1.36$ & $9.93$ &  $0.00376$  \\  \hline 
\end{tabular}
\caption{Parameter sets $A_3$, $B_3$, $C_3$ and $D_3$ corresponding to the representative tumours in the BS cohort.}
\label{tab:4.3}
\end{wraptable}

More specifically, tumours $A_3$ and $B_3$ respond to RT similarly to SL tumours $C_2$ and $D_2$, respectively (recall Figures \ref{fig:4.2.7} and \ref{fig:4.2.8}), while tumour $D_3$ responds similarly to NL tumour $D_1$ (recall Figures \ref{fig:4.2.4} and \ref{fig:4.2.5}).  For tumour $A_3$,  this involves an initial large increase in viable and total volume as the tumour evolves towards its SL steady state, followed by a substantial increase in RT cell kill, the average oxygen concentration and the dead cell volume.  Despite the reduction in viable volume, the accumulation of dead material implies a significant increase in total volume. The same qualitative behaviour is observed for tumour $B_3$,  with less RT-induced cell death and dead material accumulation as the oxygen concentration remains significantly lower than for $A_3$.  As a result, the increase in total volume is also smaller. For tumour $D_3$,  cell death due to RT and hypoxia is outweighed by rapid tumour regrowth between fractions, leading to small increases in viable and total volumes.

Further, while tumour $C_3$ lies closest to the SL region ($V_d \approx 0.8$), it does not transition from the basin of attraction of its NL steady state to its SL steady state, unlike tumours $A_3$ and $B_3$.  In particular, the increase in the oxygen concentration for $C_3$ is not rapid enough for the tumour to enter, during treatment, a SL regime where, on average, $c > c_{min}$.  Therefore, the increase in viable volume is constant, but gradual, with a smaller accumulation of dead material. This explains why $C_3$ undergoes a larger than average increase in viable volume, with a moderate increase in total volume.

In summary, we have identified two extremal regions of parameter space in which tumours in a BS regime undergo larger decreases or increases in viable volume (and larger increases in total volume) than the typical tumour in this regime. Tumours which are sufficiently near to the boundary of the BS and SL regimes and consume little oxygen for maintenance experience larger than average decreases in viable cell volume as RT cell death is enhanced by higher oxygen levels. By contrast, tumours which are close to the boundary between the BS and SL regimes, but not sufficiently close, undergo larger than average increases in viable volume, regardless of the value of $q_1$.  This occurs as they attempt and fail to transition from their NL steady state to their SL steady state and, thus, RT cell death remains limited by low oxygen levels and outweighed by tumour regrowth between fractions.

\begin{figure}[!h]
\centering
\includegraphics[scale=0.4]{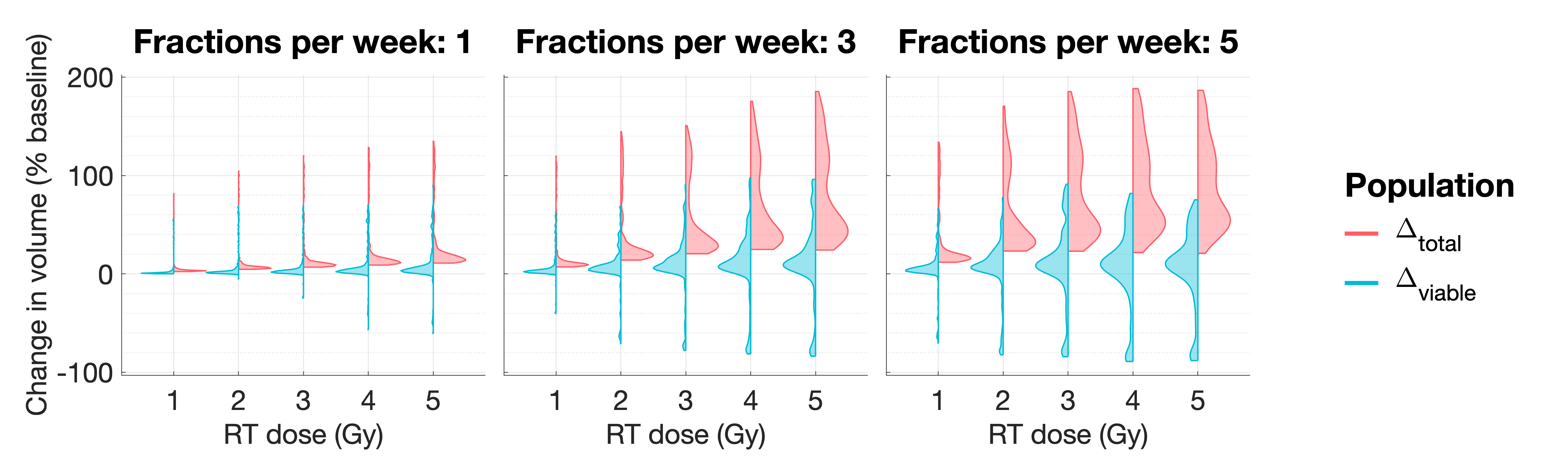} 
\caption{For the virtual BS population, we show how the distributions of $\Delta_{viable}$ and $\Delta_{total}$ change as the dose rate, $R \in \llbracket 0.1, 0.5 \rrbracket$, and the number of fraction per week, $N_{frac} \in \{1,3,5\}$, vary. The mean and maximum values of $\Delta_{viable}$ increase, while its minimum value decreases, as $R$ and $N_{frac}$ increase. The mean, minimum and maximum values of $\Delta_{total}$ increase as $R$ and $N_{frac}$ increase. There is an exception for $N_{frac}=5$, where the maximum value of $\Delta_{viable}$ and minimum value of $\Delta_{total}$ decrease with $R \geq 3$. }
\label{fig:4.2.16}
\end{figure}

\paragraph{The effect of the dosing schedule on typical tumour response.}
For the virtual population of tumours in a BS regime, we show in Figure \ref{fig:4.2.16} how the dose rate, $R$, and the number of fractions per week, $N_{frac}$, affect tumour response to RT when the total dose is fixed. 
On average, a higher number of fractions per week (for fixed $R$) and a higher dose rate (for fixed $N_{frac}$) lead to greater increases in the viable and total cell populations.  While these results contrast with those for tumours in SL regimes, we see that the maximum reduction in viable volume increases with $R$ and $N_{frac}$, similarly to SL tumours. Overall, these results indicate that, in most cases, increasing the RT dose and frequency may be deleterious (similarly to NL tumours).

\subsection{Post-treatment tumour growth dynamics}
In the previous section, we discussed the short-term response to RT of tumours in different growth regimes, distinguishing between tumours in monostable (NL and SL) and bistable regimes.  We now investigate the long-term response to RT by studying post-treatment tumour growth dynamics and, in particular, the tumour steady states attained following treatment.
\subsubsection{Steady state analysis}
We first perform a steady state analysis of the system \eqref{eq:2.7}-\eqref{eq:2.10} to understand the potential long-term effects of RT. Upon completion of a radiation protocol, we have $R \equiv 0$ thereafter. We, therefore, seek steady state solutions by setting $R=0$ and $\frac{\mathrm{d}}{\mathrm{d}t} = 0$ in Equations \eqref{eq:2.7}-\eqref{eq:2.10} and solving the following system
\begin{equation}
 q_2 cT(1 - \Sigma) - \delta_1(c_{min} - c)H(c_{min}-c)T + \mu T_{S} = 0,
 \label{eq:4.3.1}
\end{equation}
\begin{equation}
 {\theta_2 q_2 }cT_{S}(1 - \Sigma) - (\delta_{1,S}(c_{min} - c)H(c_{min}-c) + \mu +\xi ))T_S = 0,
 \label{eq:4.3.2}
\end{equation}
\begin{equation}
\xi T_{S} - \eta_R T_{R} = 0,
\label{eq:4.3.3}
\end{equation}
\begin{equation}
g(1-c)V_0 - q_1 (T + \theta_1 T_S) c  - q_3 \left(T+\theta_2 T_S\right)c(1- \Sigma)= 0.
\label{eq:4.3.4} 
\end{equation}

Denoting the steady state solutions by $T^*$, $T_S^*$, $T_R^*$ and $c^*$, respectively, Equation \eqref{eq:4.3.3} implies that $T_S^* = \frac{\eta_R}{\xi} T_R^*$. Therefore, we have either $T_S^* = T_R^* = 0$ or $T_S^*,\,T_R^* > 0$.  Suppose that $T_S^*,\,T_R^* > 0$.  We can show, by contradiction, that there are no physically realistic steady state solutions satisfying this condition by, first, proving that there are no SL steady states with $T_S^*,\,T_R^* > 0$ and, then, proving that there are no NL steady states with $T_S^*,\,T_R^* > 0$.  

If $c^* \geq c_{min}$, then Equation \eqref{eq:4.3.1} gives
\begin{equation}
T^* = \frac{- \mu T^*_{S}}{q_2 c^*(1 - \Sigma^*_R)} < 0,
\label{eq:4.3.5}
\end{equation}
since $T_S^* > 0$ and $\Sigma^* < 1$ by assumption and $\mu > 0$ and $q_2 > 0$ by definition.  Since $T^* >0$ is required for a physically realistic solution, there are no SL steady states with $T_S^*,\,T_R^* > 0$.

If $ 0 < c^* < c_{min}$,  then Equation \eqref{eq:4.3.1} supplies
\begin{equation}
\frac{ q_2 }{\delta_1} (1 - \Sigma^*) - \frac{(c_{min} - c^*)}{c^*} = - \frac{\mu T^*_S}{\delta_1 c^* T^*} < 0.
 \label{eq:4.3.6}
\end{equation}

Since $q_2 = \delta_1$, we have
\begin{equation}
(1 - \Sigma^*) - \frac{(c_{min} - c^*)}{c^*} = - \frac{\mu T^*_S}{\delta_1 c^* T^*} < 0.
 \label{eq:4.3.7}
\end{equation}

Then, Equation \eqref{eq:4.3.2} implies that
\begin{equation}
{\frac{\theta_2 q_2}{ \delta_{1,S}} (1 - \Sigma^*) -\frac{(c_{min}-c^*)}{c^*} = \frac{(\mu +\xi)}{\delta_{1,S}} c^*} > 0,
\label{eq:4.3.8}
\end{equation}

Since $\theta_2 q_2 = \delta_{1,S}$, we have
\begin{equation}
(1 - \Sigma^*) - \frac{(c_{min}-c^*)}{c^*} = \frac{(\mu +\xi)}{\delta_{1,S}} c^* > 0
 \label{eq:4.3.9}
\end{equation}

Comparing Equations \eqref{eq:4.3.7} and \eqref{eq:4.3.9},  we obtain a contradiction. This implies that there are no NL steady states with $T_S^*,\,T_R^* > 0$.  We, therefore, conclude that NL and SL steady state solutions of the system \eqref{eq:2.7}-\eqref{eq:2.10} must have $T_S^*= T_R^* = 0$. It is then straightforward to show that the solutions of the system \eqref{eq:4.3.1}-\eqref{eq:4.3.4} with $T_S^*= T_R^* = 0$ are equal to the steady state solutions in the absence of treatment \citep{colson2022combining} (see Appendix A). 

We have shown that RT preserves the steady states and growth regimes observed in the absence of treatment.  We conclude that, given $T(0) = T^*$, tumours in monostable regimes at the start of treatment will return to their original tumour volume, $\Sigma_0 = T^* + V_0$, and composition ($T_S^* = T_R^* = 0$) after RT. In contrast, tumours in a BS regime either return to the original, NL steady state or evolve to the SL steady state. 

\subsubsection{Elucidating conditions for RT to drive steady state switching of tumours in bistable regimes} \label{sec:4.3.2}
\begin{figure}[!h]
\centering
\includegraphics[scale=0.45]{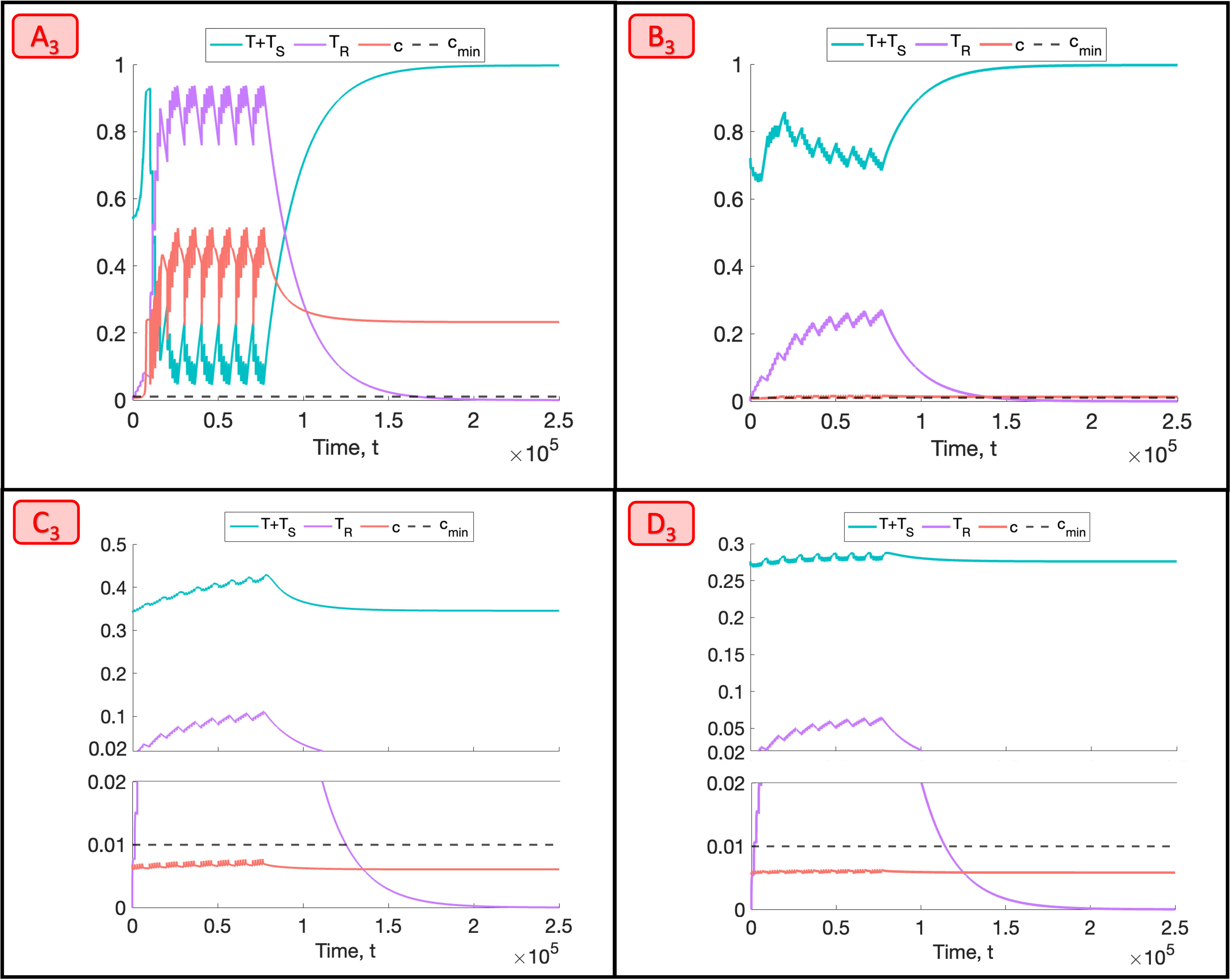} 
\caption{For a conventional fractionation schedule, we numerically solve Equations \eqref{eq:2.7}-\eqref{eq:2.10} for $t \in (0,2.5 \times 10^5]$ subject to the initial conditions \eqref{eq:4.1.3}. In $A_3$-$D_3$, we fix $V_0 = 0.00275$ and $(q_1,q_3)$ as indicated by the points $A_3$, $B_3$, $C_3$ and $D_3$ in Figure \ref{fig:4.2.11}, which corresponds to tumours in a BS regime.  Tumours $C_3$ and $D_3$ evolve to their NL steady states following treatment, whereas tumours $A_3$ and $B_3$ switch to their SL steady state following treatment.}
\label{fig:4.3.1}
\end{figure}

The steady state analysis showed that tumours in a BS regime may attain either a NL or a SL steady state following treatment. In particular, such tumours may undergo large increases in tumour volume in response to RT as they switch from a NL steady state to a larger SL steady state.  Recall the tumours $A_3$-$D_3$ that we defined in Section \ref{sec:4.2.2}: Figure \ref{fig:4.3.1} shows their response to RT both during and following a conventional fractionation schedule.

Tumours $C_3$ and $D_3$ underwent increases in viable volume during treatment and then returned to their NL steady state following treatment: the effect of RT was not strong enough to cause a switch in steady state. By contrast, tumours $A_3$ and $B_3$ experienced reductions in viable volume during treatment and then evolved to their SL steady state following treatment.  The oxygen concentration in both of these tumours increased beyond the hypoxic threshold, $c_{min}$, during ($A_3$) or following ($B_3$) RT and remained above this threshold level thereafter. This enabled the viable cell population to grow unchecked until the SL equilibrium was reached.

In contrast to tumours $C_3$ and $D_3$, we recall that tumours $A_3$ and $B_3$ are characterised by $V_d \approx 1$, where $V_d$ is defined in \eqref{eq:4.2.1}.  They are also, respectively, characterised by high and low values of $q_1$, the oxygen consumption rate for maintenance. This suggests that tumours which are near to the boundary between BS and SL regions in parameter space are most susceptible to undergoing a switch in steady state volume in response to RT, irrespective of the value of $q_1$. This observation holds across a range of RT protocols (see Appendix B).

\begin{figure}[!h]
\begin{subfigure}[t]{0.3\textwidth}
\centering
\includegraphics[scale=0.43]{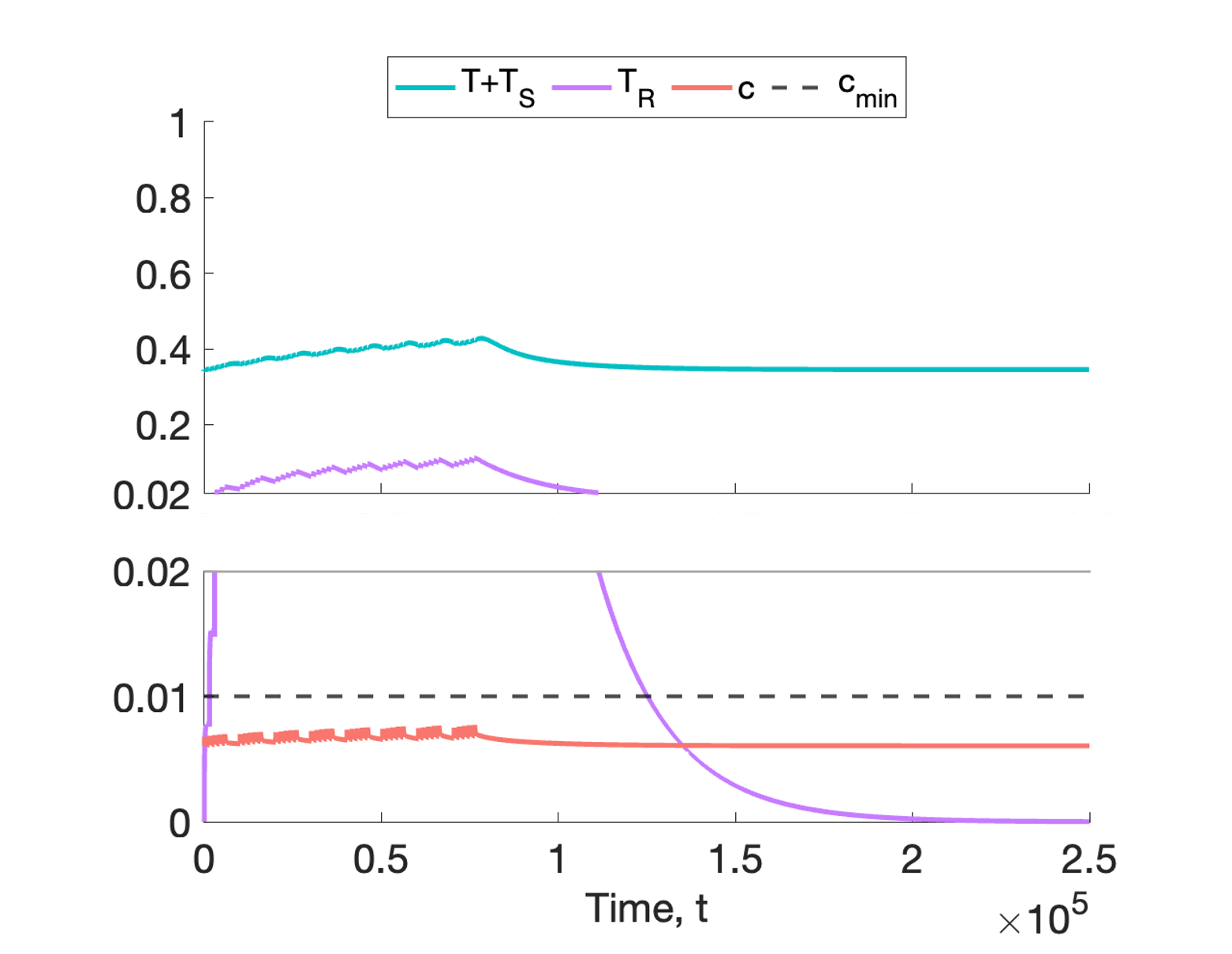} 
\caption{}
\label{fig:4.3.2A}
\end{subfigure}
\hfill
\begin{subfigure}[t]{0.3\textwidth}
\centering
\includegraphics[scale=0.43]{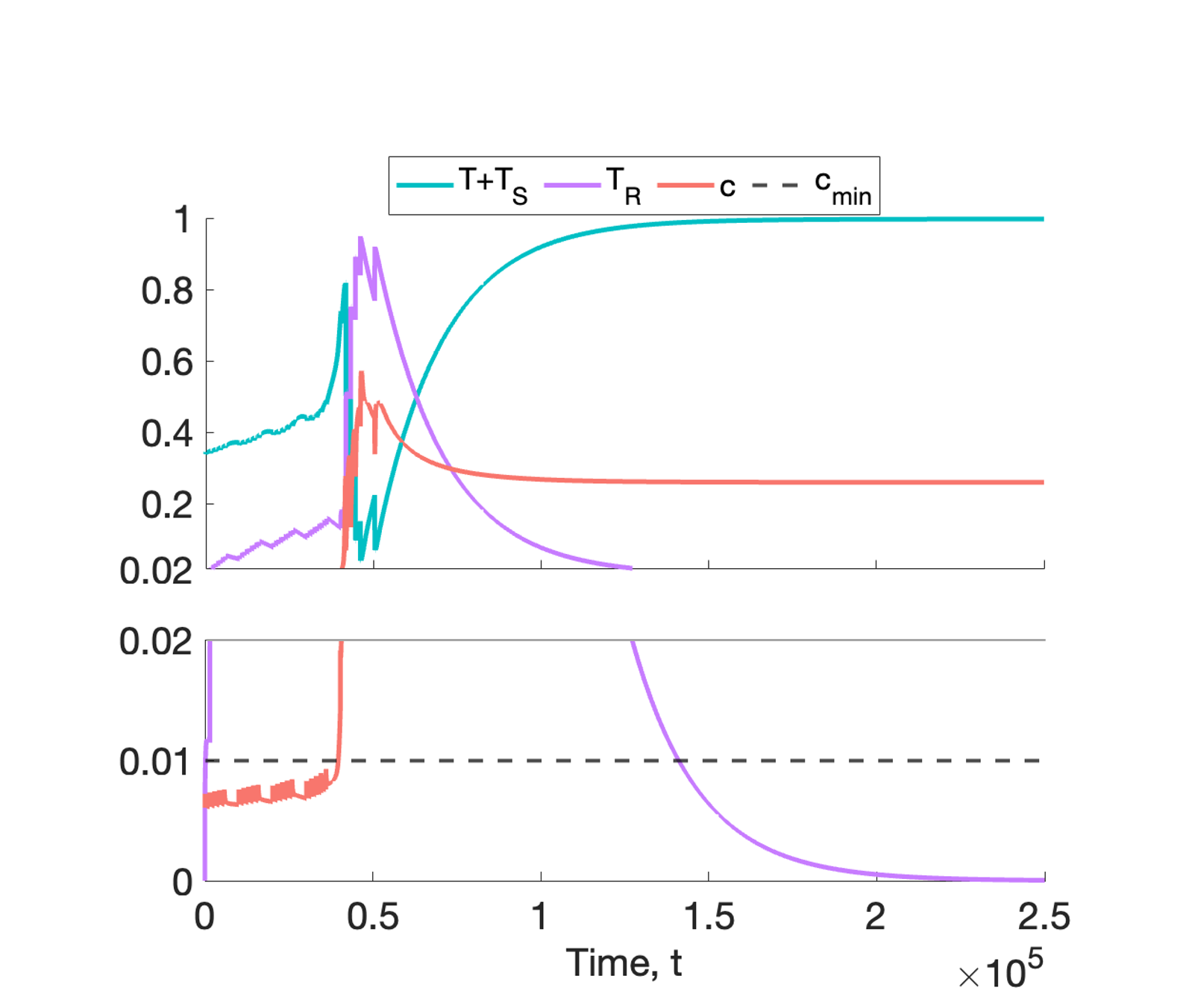} 
\caption{}
\label{fig:4.3.2B}
\end{subfigure}
\hfill
\begin{subfigure}[t]{0.3\textwidth}
\centering
\includegraphics[scale=0.43]{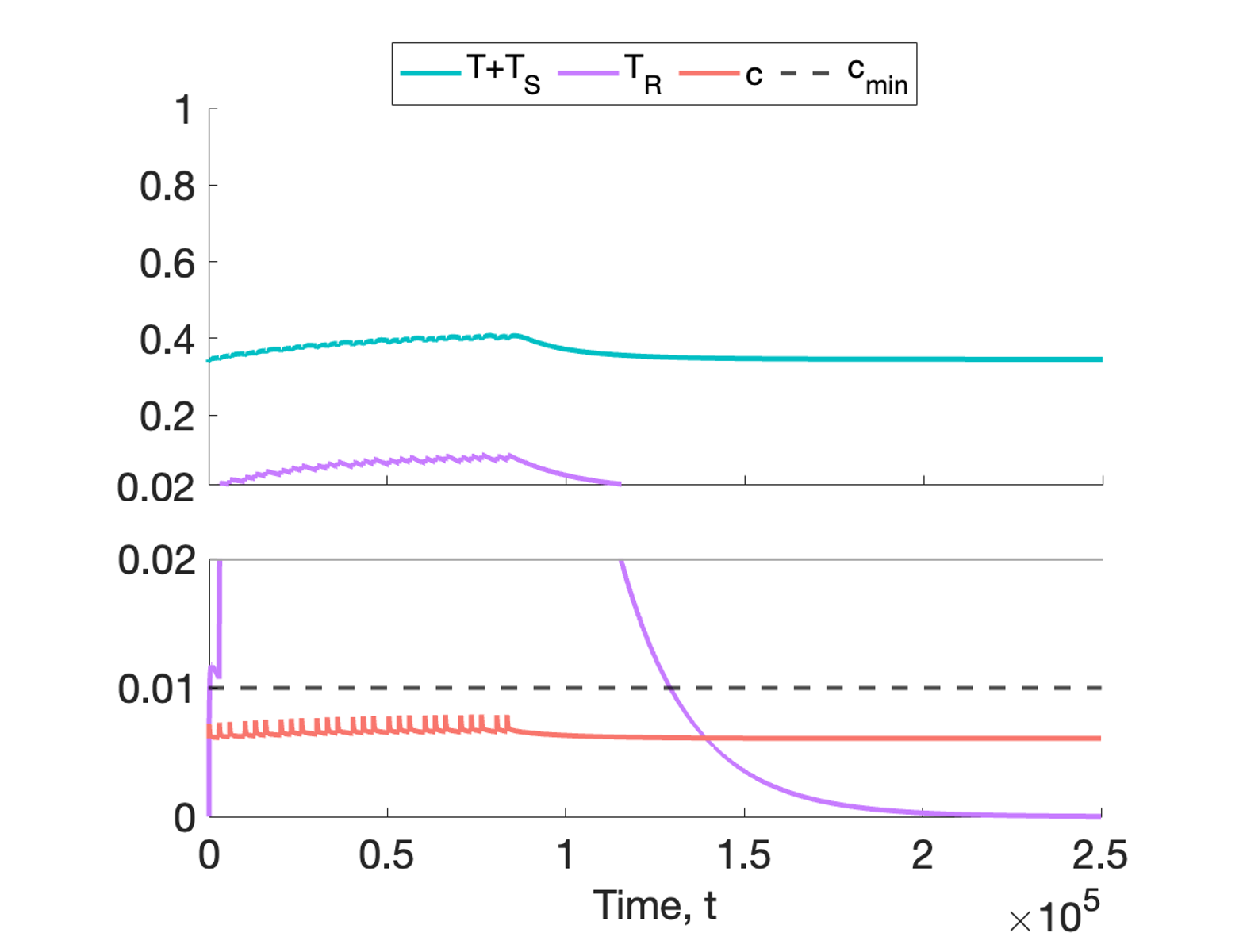} 
\caption{}
\label{fig:4.3.2C}
\end{subfigure}
\caption{We numerically solve Equations \eqref{eq:2.7}-\eqref{eq:2.10} for $t \in (0,2.5 \times 10^5]$ subject to the initial conditions \eqref{eq:4.1.3}. We impose the dose rates (a) $R = 0.2$ and (b),(c) $R = 0.3$ and simulate (a) daily fractions, Monday to Friday, for $8$ weeks, (b) daily fractions, Monday to Friday, for $5.2$ weeks and (c) fractions on Monday, Wednesday and Friday for $8.67$ weeks. We fix $V_0 = 0.00275$ and $(q_1,q_3)$ as indicated by the parameter set $C_3$ in Table \ref{tab:4.3}. Comparing (a) and (b) and (b) and (c) shows how lower RT doses and less frequent dosing both prevent the tumour $C_3$ from evolving to the SL steady state following treatment. }
\label{fig:4.3.2}
\end{figure}

We now consider how the dosing schedule affects the long-term dynamics of tumours in a BS regime.  In Figures \ref{fig:4.3.2A} and \ref{fig:4.3.2B}, we show the response of tumour $C_3$ to two fractionation protocols comprising either $2$ or $3$ Gy fractions applied 5 times per week for $8$ or $5.2$ weeks, respectively.  A switch in steady state is observed for $3$ Gy fractions. This suggests that the likelihood of a tumour switching steady state increases with dose,  a consistent trend in our numerical study (see Appendix B). Figure \ref{fig:4.3.2C} additionally shows the response of tumour $C_3$ to $3$ Gy fractions applied $3$ times per week for $8.67$ weeks. Comparing this figure to Figure \ref{fig:4.3.2B} highlights how a lower dosing frequency can prevent the transition from NL to SL steady state for tumours in BS regimes (see Appendix B). 

These results suggest that a lower RT dose and dosing frequency may prevent uncontrolled increases in tumour volume following RT for tumours in BS regimes. As with our observations for short-term treatment responses, this challenges the assumption that a higher dose, applied with a higher frequency, will lead to a greater reduction in tumour volume.

\section{Discussion} \label{sec:5}
Cancer is a heterogeneous disease. In particular,  tumours can exhibit widely varying responses to treatments.  As a result, the success of existing therapies, which are typically applied following a ``one-size-fits-all approach'', can be highly variable.  Patient-specific treatment design could aid in overcoming these barriers to treatment efficacy,  but this requires increased understanding of the factors which affect tumour sensitivity to treatment. In this paper, we investigated how two distinct mechanisms of growth arrest can influence tumour responses to radiotherapy (RT). 

We extended an existing model of tumour growth which distinguishes between nutrient limited (NL) and space limited (SL) growth control \citep{colson2022combining}. In the absence of treatment, this model exhibits three growth regimes: (i) NL, where a tumour attains a NL steady state at which cell proliferation and death balance, (ii) SL, where a tumour attains a SL steady state when cell proliferation ceases due to space constraints, with no cell death, and (iii) bistable (BS), where stable NL and SL steady states coexist. In this paper, we investigated how tumours in each regime respond to RT.  We found that the short- and long-term responses of tumours in monostable regimes (i.e. NL and SL) can be distinguished from that of tumours in BS regimes.

Tumours in the SL regime typically respond well to RT in the short-term, as both their viable and total volumes decrease during fractionation, while tumours in the NL regime typically respond less well, since their total volume increases despite a reduction in viable volume. However, certain NL and SL tumours respond significantly better than the average tumour in their respective regimes.  By identifying parameter regions which give rise to these outliers, we determined different mechanisms that underpin successful RT. For NL tumours, RT efficacy is maximised when regrowth between fractions is minimised, while, for SL tumours, increased RT efficacy may be due to limited regrowth (as for NL tumours) and/or RT cell kill.  The additional SL-specific mechanism is a consequence of low rates of RT cell kill for NL tumours due to low oxygenation. This explains how the different growth arrest mechanisms that characterise the NL and SL regimes can affect short-term tumour response to RT. In the long-term, tumours in NL and SL regimes always return to their pre-treatment steady state volume, irrespective of the effects of RT during treatment.  Our model therefore predicts that any change in the tumour burden during radiation is transient for these tumours. 

We also found that most tumours in a BS regime respond badly to RT in the short-term, as their viable and total cell volumes increase during RT.  As for monostable regimes, outliers which lie, in parameter space, near the boundary between BS and SL regions, exhibit more extreme responses to RT. In these cases, the intratumoural oxygen concentration is close to and smaller than $c_{min}$, the threshold concentration below which cells die due to nutrient insufficiency.  If RT induces a net increase in oxygen levels such that $c > c_{min}$,  cell death due to nutrient insufficiency ceases and RT drives the tumour to its SL steady state.  This leads to a significant increase in RT-induced cell death and dead cell accumulation, resulting in large decreases and increases in viable and total volumes, respectively.  By contrast, if RT induces a net increase in oxygen levels such that $c \leq c_{min}$,  RT causes large increases in viable and total volumes as the tumour grows towards, and fails to reach, its SL steady state. Here, RT cell kill is outweighed by tumour growth between fractions throughout treatment.  Irrespective of whether these outliers experience increases or decreases in viable volume, they evolve to their larger SL steady state following RT.  Therefore, the model predicts that, in a BS regime, RT usually has a detrimental effect on tumour growth.

A final key result relates to RT dosing schedules. We found that, in SL regimes, applying larger doses at higher frequency typically increases RT efficacy, whereas, in NL and BS regimes, administering lower doses at lower frequency can increase RT efficacy for outliers and lessen or prevent large increases in tumour burden across the virtual cohorts. The latter is a counter-intuitive result and challenges the assumption that giving the maximum tolerable dose is the best course of treatment. In practice, we are unlikely to know which growth regime a patient's tumour lies in when treatment starts. It would be interesting, in future work, to investigate whether we can determine a tumour's growth regime by monitoring its response to a given treatment protocol.  If we can establish that a tumour is in a SL regime, this would allow us to adapt the treatment protocol to maximise the reduction in tumour burden, e.g., by increasing the RT dose or dosing frequency. Alternatively, if a tumour is in a NL or BS regime, it might be preferable to halt treatment early in order to prevent large increases in tumour burden.  

In this paper, we studied the effects of RT on tumour cells and neglected its effects on the tumour vasculature.  In particular, we viewed the vascular volume as a parameter which influences a tumour's carrying capacity, rather than a dynamic variable. This simplifying assumption will cease to be valid at long times when effects such as angiogenesis and vascular remodelling become important. In future work, we will extend our model to relax this assumption, and obtain a more realistic description of tumour response to treatments which affect both tumour and endothelial cells. 

\appendix

\section{Steady state solutions in the absence of treatment}
A steady state analysis of the system \eqref{eq:2.7}-\eqref{eq:2.10} with $R \equiv 0$ was performed in \citep{colson2022combining}. There exist two SL steady states given by
\begin{align}
    \mathrm{SS}_1:& \quad (T_1^*,c_1^*)  = (0,1), \label{eq:A.1}\\
    \mathrm{SS}_2:& \quad(T_2^*,c_2^*)  =  \left(1 - V_0, \frac{V_0}{V_0 + (q_1/g) (1-V_0)}\right).\label{eq:A.2}
\end{align} 
$ \mathrm{SS}_1$ is unstable for all combinations of parameters, while $ \mathrm{SS}_2$ is stable in the parameter regions in which it is an admissible solution. 

There are also two NL steady states given by
\begin{align}
    \mathrm{SS}_3:& \quad (T_3^*,c_3^*)  = \left(T(c_-), c_-\right) \label{eq:A.3}\\[5pt]
    \mathrm{SS}_4:& \quad(T_4^*,c_4^*)  = \left(T(c_+), c_+\right), \quad \text{if } V_0  \neq \frac{2(q_3-q_1)}{g + q_3-q_1},\label{eq:A.4}
\end{align}
where
\begin{equation}
T(c) = (1 - V_0) - \left(\frac{c_{min}}{c} -1\right),
\label{eq:A.5}
\end{equation}
and
\begin{subnumcases}{}
 {c_{\pm}  =  \frac{c_{min}(X \mp \sqrt{X^2 + 4 q_3 Y})}{2Y},  \quad  \qquad \qquad \, \qquad \quad \,\, \text{if } V_0  \neq \frac{2(q_3-q_1)}{g + q_3-q_1},}  \label{eq:A.6}
\\
{c_- = \frac{c_{min}(q_3(g + q_3-q_1))}{2\frac{g}{c_{min}}(q_1-q_3)-g(q_1-3q_3)+(q_1-q_3)^2}, \quad \,\, \text{if } V_0  = \frac{2 (q_3-q_1)}{g+q_3-q_1}},  \label{eq:A.7}
\end{subnumcases}
with
\begin{equation}
\begin{cases}
X = q_1-3q_3+\left(\frac{g}{c_{min}}+q_3\right)V_0,\\[8pt]
Y = 2(q_1-q_3)+(g+q_3 -q_1)V_0.
\end{cases}
\label{eq:A.8}
\end{equation}
In the regions in which $ \mathrm{SS}_3$ and  $ \mathrm{SS}_4$ exist as admissible steady state solutions, $ \mathrm{SS}_3$ is stable, while $ \mathrm{SS}_4$ is unstable.

\section{Numerical results: steady state switching of tumours in the bistable regime}

\begin{figure}[!h]
    \centering
    \includegraphics[scale = 0.5]{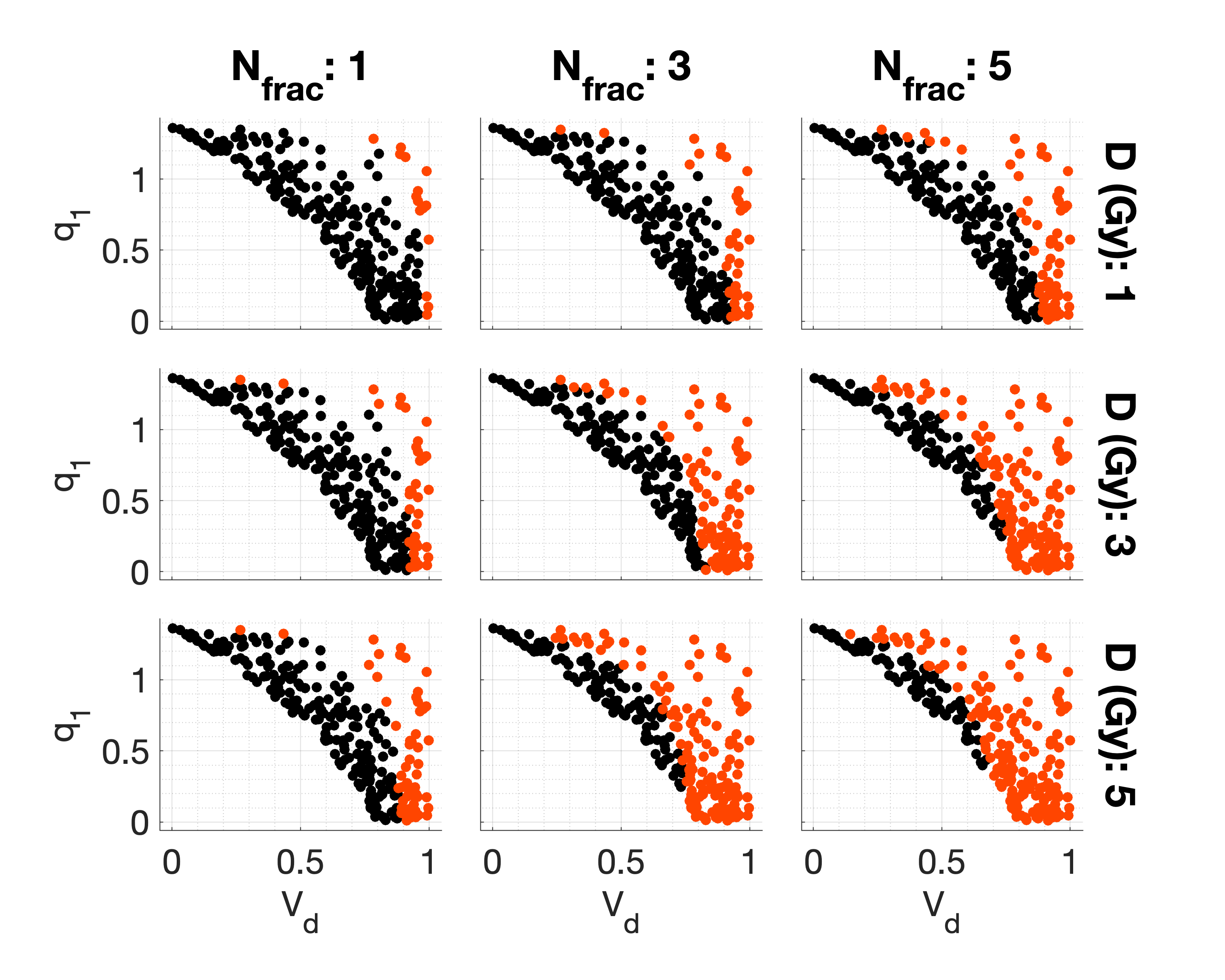}
    \caption{For the virtual BS tumour population and fractionation schedules with $R \in \{ 0.1, 0.3, 0.5 \} $ and $N_{frac} \in \{1,3,5\}$, the scatter plots show the $(q_1,V_d)$ pairs that correspond to tumours that switch (red) and do not switch (black) steady state. The former are typically characterised by larger values of $V_d$.}
    \label{fig:A.1}
\end{figure}

For dosing regimens with $R \in \{0.1,0.3,0.5 \}$ and $N_{frac} \in \{1, 3, 5 \}$, the scatter plots in Figure \ref{fig:A.1} highlight (in red) the $(q_1,V_d)$ pairs which correspond to tumours in a bistable regime that switch steady state. We observe that tumours which switch steady state typically have larger values of $V_d$. We note also that the number of tumours which switch steady state increases with the $R$ and $N_{frac}$. These results are consistent with those presented in Section \ref{sec:4.3.2} for tumour $C_3$. 

\bibliography{Manuscript_bib}

\end{document}